\documentclass[final,5p,times,twocolumn]{elsarticle}

\usepackage{amssymb}
\usepackage{amsmath}
\usepackage{amsthm}

\usepackage{times}
\usepackage{epsfig}
\usepackage{graphicx}
\usepackage{tabularx}
\usepackage{booktabs}
\usepackage{multirow}
\usepackage{makecell}
\usepackage{xcolor} 

\newcommand{\revise}[1]{\textcolor{black}{#1}}
\newcommand{\reviseminor}[1]{\textcolor{black}{#1}}
\newcommand{\et}[2]{${#1}^{\pm{#2}}$}
\newcommand{\etb}[2]{$\mathbf{{#1}}^{\pm{#2}}$}
\newcommand{\ets}[2]{$\underline{{#1}}^{\pm{#2}}$}

\usepackage{etoolbox}
\newcommand{\etal}{\textit{et al.}}

\journal{Graphical Models}

\begin{document}

\begin{frontmatter}



\title{Semantics-Aware Human Motion Generation from Audio Instructions} %% Article title


\author[UCAS,siat]{Zi-An Wang} %% Author name
\author[siat]{Shihao Zou}
\author[SUST,siat]{Shiyao Yu}
\author[ntu]{Mingyuan Zhang}
\author[siat]{Chao Dong}

\affiliation[UCAS]{organization={University of Chinese Academy of Sciences},%Department and Organization
            % city={Shenzhen},
            country={China}}

\affiliation[siat]{organization={Shenzhen Institutes of Advanced Technology, Chinese Academy of Sciences},%Department and Organization
            % city={Shenzhen},
            country={China}}

\affiliation[SUST]{organization={Southern University of Science and Technology},%Department and Organization
            % city={Shenzhen},
            country={China}}

\affiliation[ntu]{organization={Nanyang Technological University},%Department and Organization
            % city={Shenzhen},
            country={Singapore}}

%% Abstract
\begin{abstract}
% Recent advancements in interactive technologies, such as GPT-4o, have highlighted the growing prominence of using audio signals to encode semantics. Building on this trend, this paper explores a relatively new task: human motion generation from audio instructions, where audio signals are used as conditioning inputs to generate motions that align with the semantics of the audio. Compared to text-based user interactions, audio provides a more convenient and natural mode of communication in real-world scenarios. However, existing audio-conditioned methods primarily focus on aligning movements with the rhythm of music or speech, which often results in a weak connection between the semantics of the audio and the generated motions.
% In this work, we propose an end-to-end framework based on masked generative transformer, incorporating a memory-retrieval based attention module to address the challenges posed by sparse and lengthy audio signals. Additionally, we augment existing datasets by rephrasing textual descriptions into a more conversational, spoken language style, and synthesizing corresponding audio with various speaker identities. Experimental results demonstrate both the effectiveness and efficiency of our framework for generating human motion from audio instructions. Notably, audio exhibits a comparable ability to text in representing semantics, while offering more efficient and user-friendly interactions in practical applications.

Recent advances in interactive technologies have highlighted the prominence of audio signals for semantic encoding. This paper explores a new task, where audio signals are used as conditioning inputs to generate motions that align with the semantics of the audio. Unlike text-based interactions, audio provides a more natural and intuitive communication method. However, existing methods typically focus on matching motions with music or speech rhythms, which often results in a weak connection between the semantics of the audio and generated motions. We propose an end-to-end framework using a masked generative transformer, enhanced by a memory-retrieval attention module to handle sparse and lengthy audio inputs. Additionally, we enrich existing datasets by converting descriptions into conversational style and generating corresponding audio with varied speaker identities. Experiments demonstrate the effectiveness and efficiency of the proposed framework, demonstrating that audio instructions can convey semantics similar to text while providing more practical and user-friendly interactions.
\end{abstract}


\begin{graphicalabstract}
\includegraphics[width=\linewidth]{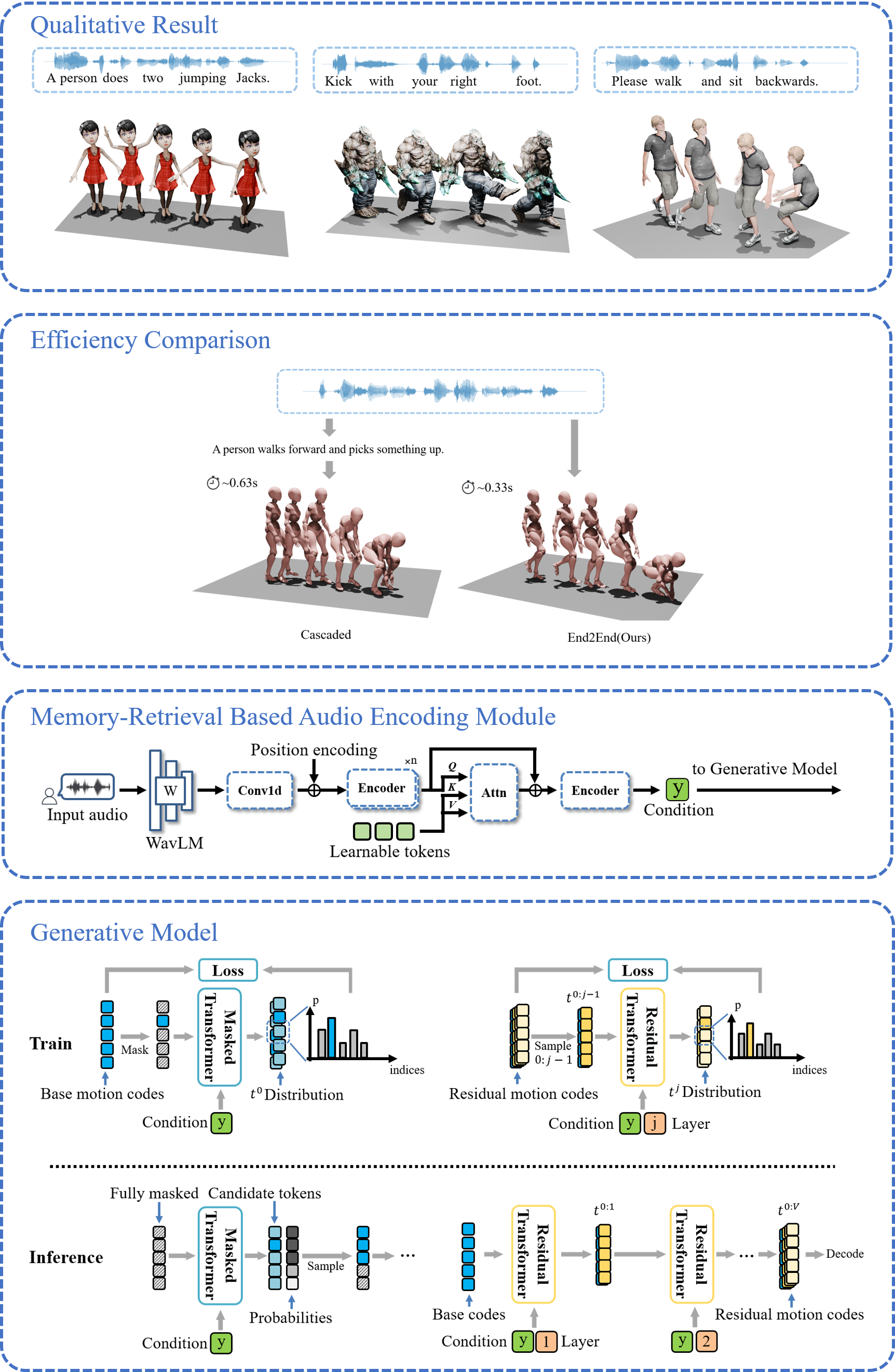}
\end{graphicalabstract}

\onecolumn
\begin{highlights}
\item We introduce a relatively new task in this paper: human motion generation from audio instructions for a more convenient user interaction system. To achieve this, we design an end-to-end framework based on the masked generative transformer, incorporating a memory-retrieval based attention module to handle the challenges posed by sparse and lengthy audio signals.
\item We also augment existing text-based motion generation datasets for this task by rephrasing textual descriptions into a more conversational, spoken language style, and synthesizing corresponding audio with various speaker identities.
\item The experimental results demonstrate the effectiveness and efficiency of our proposed end-to-end framework for generating human motion directly from audio instructions. Notably, the findings indicate that audio offers a comparable ability to text in representing semantics, making it a strong alternative for conditioning signals in developing more efficient and user-friendly systems.
\item We conduct the experiments on both the Original Dataset and the augmented Oral Dataset demonstrating the effectiveness of our method under audio instructions with different styles.
\vfill
\end{highlights}
\twocolumn
%% Keywords

\end{frontmatter}



\section{Introduction}
\label{sec:introduction}
% With the advancement of technologies such as the metaverse, the significance of user interaction convenience has markedly increased. Audio, serving as a direct channel for human semantic understanding, has garnered substantial attention among various modalities. Recently, numerous interactive technologies leveraging audio signals have emerged. For instance, GPT-4o~\cite{openai2024gpt} employs audio as an interaction modality, significantly enhancing user convenience and immersion. This underscores the advantage of audio in facilitating ease of interaction.

% In the domain of human motion generation, audio-conditioned tasks primarily encompass music-to-dance~\cite{chen2021choreomaster,li2021ai,dabral2023mofusion,gong2023tm2d} and speech-to-gesture~\cite{ao2023gesturediffuclip,ao2022rhythmic,li2021audio2gestures,zhang2024semantic}. However, the conditioning signals in these tasks are often inadequate for providing explicit motion descriptions~\cite{zhu2023survey}, resulting in diminished controllability of the generated motions.

\begin{figure*}[h]
    \centering
    \includegraphics[width=\linewidth]{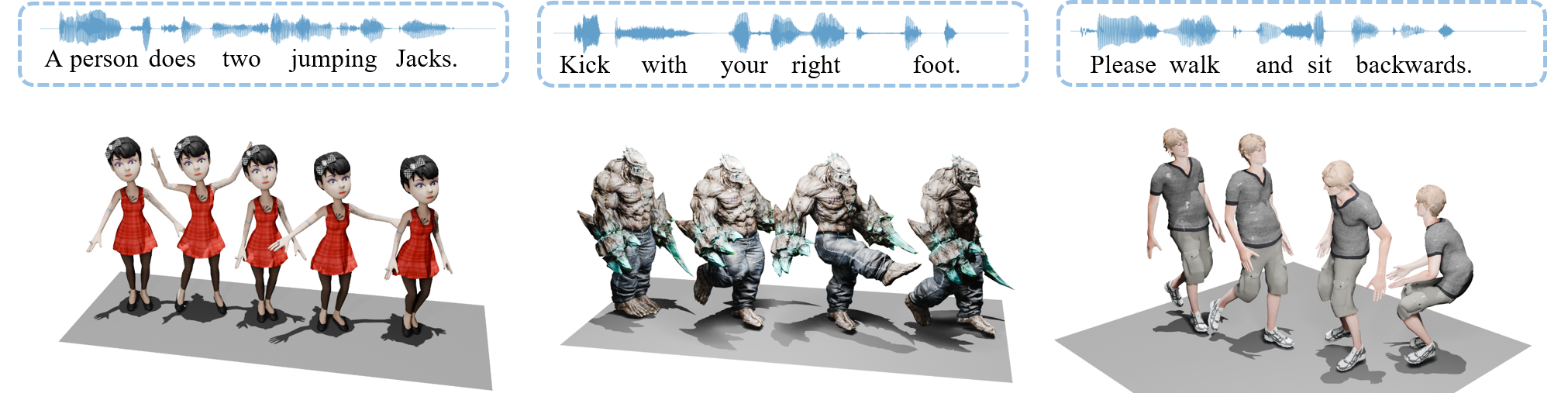}
    \caption{
    \textbf{Overview of Our Work}. Given an audio instruction as the conditional signal (with \textbf{text included for reference purposes only}), our generative model is able to produce high-quality human motion sequences that accurately align with the semantics of the audio input.
    }
    % \vspace{-10pt}
    \label{fig:audio_performance}
\end{figure*}

\revise{Language-guided human motion generation aims at generating human motion sequences aligned with the given natural language description on details like action, aim, or speed. This task has gained increasing attention~\cite{zhu2023survey} due to its wide range of applications, such as in the metaverse, video games, and virtual reality. 
\reviseminor{
Many existing works focus on text-conditioned motion generation~\cite{plappert2016kit,guo2022generating,tevet2023human,zhang2023generating,shi2024generating} by leveraging large language pre-trained models~\cite{devlin2018bert,radford2021learning} to encode text semantics.
}
However, text-based user interaction poses significant challenges in real-world applications, as it often requires users to articulate precise and detailed descriptions, which can be time-consuming and unintuitive. This complexity makes it less practical for dynamic or fast-paced environments where quick, natural, and seamless interaction is essential for user engagement.}

% Language-guided human motion generation~\cite{zhu2023survey} has gained increasing attention due to its wide range of applications, such as in the metaverse, video games, and virtual reality. Many existing works focus on text-conditioned motion generation~\cite{plappert2016kit,guo2022generating,tevet2023human,zhang2023generating} by leveraging large language pre-trained models~\cite{devlin2018bert,radford2021learning} to encode text semantics. However, text-based user interaction poses significant challenges in real-world applications, as it often requires users to articulate precise and detailed descriptions, which can be time-consuming and unintuitive. This complexity makes it less practical for dynamic or fast-paced environments where quick, natural, and seamless interaction is essential for user engagement.

\reviseminor{In contrast, audio provides a more direct and natural channel for communication and semantic understanding, offering a more intuitive and seamless interaction experience. Recent advances in interactive technologies have further emphasized the potential of audio for encoding semantics. For example, GPT-4o~\cite{openai2024gpt} incorporates audio as an interaction modality, greatly enhancing user convenience and immersion. This underscores the unique advantage of audio in improving both the ease and depth of human-computer interaction. In the context of human motion generation, audio-conditioned tasks such as music-to-dance~\cite{chen2021choreomaster,li2021ai,dabral2023mofusion,gong2023tm2d,zhou2023lets} and speech-to-gesture~\cite{ao2023gesturediffuclip,ao2022rhythmic,li2021audio2gestures,zhang2024semantic,liu2025contextual,liu2022semantic} have become prominent, where motions are generated to align with the rhythm of music or speech. Specifically, the former requires generating a corresponding dance motion based on the music audio provided by the user, ensuring that the style and rhythm align with the given music. The latter involves generating motion sequences that match the rhythm of the speech audio provided by the user. However, these conditioning signals often lack explicit motion descriptions, leading to a weak connection between semantics and movement. As a result, this limitation reduces the controllability and precision of the generated motions.}

Building on these observations, we introduce a relatively new task in this paper: human motion generation from audio instructions, where audio signals are directly used as conditions to generate motions that align with the semantics of audio instructions. A straightforward approach might involve using a speech recognition algorithm, such as~\cite{radford2023robust}, to first convert the audio into text and then apply an existing text-to-motion model, like~\cite{guo2024momask}, to generate the motion. However, such a cascaded approach often incurs high computational overhead and latency. Moreover, human brain typically interprets audio signals directly without converting them into text, suggesting that a similar direct processing pathway should be pursued for artificial systems. Therefore, to achieve a more efficient and natural interaction, an approach—capable of generating motion directly from audio instructions—is essential for advancing toward artificial general intelligence.

\revise{In this paper, we propose an end-to-end framework based on the masked generative transformer. Specifically, to address the challenges of sparse and lengthy raw audio signals, we leverage WavLM~\cite{chen2022wavlm} to encode the audio into meaningful features. To further improve efficiency, we introduce a memory-retrieval based attention module that compresses these audio features into more compact, model-friendly representations. These compact features will serve as conditioning inputs for generating corresponding motions. In the area of generative models, diffusion-based generation algorithms have achieved widespread success~\cite{tevet2023human}. However, the low efficiency of the diffusion process may limit their practical applications. So we adopt the masked generative approach~\cite{chang2022maskgit} to build the generative framework. Our generative framework comprises three key components, inspired by~\cite{chang2023muse,guo2024momask}: Residual VQ (RVQ), Masked Transformer, and Residual Transformer. RVQ performs multi-layer residual quantization on motion sequences, converting them into multi-level latent motion tokens. The Masked Transformer, conditioned on the encoded audio features, leverages the masked generative paradigm~\cite{chang2022maskgit} to generate the base-level quantization codes for these tokens. Then the Residual Transformer refines the motion representation by iteratively generating residual-level quantization codes, layer by layer, progressively adding more detailed information. Finally, the base and residual codes are combined and passed through the RVQ motion decoder, which reconstructs the complete motion sequence. This multi-step process ensures that both the broad structure and finer details of the motion are captured, resulting in more accurate and realistic motion generation.}

% To address the current paucity of datasets with paired instructive audio and motion, we augment the existing large-scale text-motion paired datasets HumanML3D~\cite{guo2022generating} and KIT~\cite{plappert2016kit}. By utilizing ChatGPT~\cite{openai2024gpt}, we paraphrase text descriptions into conversational style and employ the speech synthesis algorithm Tortoise~\cite{betker2023better} to generate corresponding audio files, resulting in two versions of the dataset: base and oral.

Another challenge is the lack of datasets that pair audio instructions with corresponding human motions. Existing large-scale datasets, such as KIT~\cite{plappert2016kit} and HumanML3D~\cite{guo2022generating}, only provide paired text and motion data, denoted as \textit{original datasets}. Moreover, the style of the textual descriptions in these datasets does not reflect the natural expression of oral instructions. To address these limitations, we develop a custom prompt for ChatGPT~\cite{openai2024gpt} to rephrase text descriptions in~\cite{plappert2016kit,guo2022generating} into a more conversational, oral language style. We then use the speech synthesis tool Tortoise~\cite{betker2023better} to synthesize corresponding audios with various speaker identities. This process allows us to create augmented datasets, denoted as \textit{oral datasets}, tailored for the task of human motion generation from audio instructions.

We evaluate our proposed method on both original and oral datasets, comparing it to the latest text-based approaches using a variety of metrics. The results show that our audio-based method performs competitively with text-based methods, demonstrating the effectiveness of audio signals for semantic understanding in practical applications. Furthermore, our end-to-end generative framework is over 50\% faster than the cascaded approach, highlighting the importance of directly encoding audio semantics for a more efficient, user-friendly system. Additionally, models trained on the oral dataset significantly outperform those trained on the original dataset in audio-instructed motion generation, showcasing the value of our newly introduced dataset.

% We evaluate the proposed model on the synthesized audio-motion paired datasets based on HumanML3D and KIT. The results, assessed across multiple metrics, demonstrate that the audio-based human motion generation approach can achieve high-quality results, substantiating the capability of audio signals to accurately control human motion. Furthermore, we compare our proposed end-to-end learning algorithm with the baseline method, illustrating that our approach is more efficient.

In summary, the primary contributions of this paper are as follows:
\begin{itemize}
    \item We introduce a relatively new task in this paper: human motion generation from audio instructions for a more convenient user interaction system. To achieve this, we design an end-to-end framework based on the masked generative transformer, incorporating a memory-retrieval based attention module to handle the challenges posed by sparse and lengthy audio signals.
    \item We also augment existing text-based motion generation datasets for this task by rephrasing textual descriptions into a more conversational, spoken language style, and synthesizing corresponding audio with various speaker identities.
    \item The experimental results demonstrate the effectiveness and efficiency of our proposed end-to-end framework for generating human motion directly from audio instructions. Notably, the findings indicate that audio offers a comparable ability to text in representing semantics, making it a strong alternative for conditioning signals in developing more efficient and user-friendly systems.
\end{itemize}

\section{Related Work}

\textbf{Audio-based Motion Generation} has seen significant progress, with tasks such as music-to-dance and speech-to-gesture gaining popularity. The music-to-dance task aims to generate dance sequences synchronized with musical beats and styles. Tang \etal~\cite{tang2018dance} framed this problem as a sequence-to-sequence task, employing an LSTM~\cite{graves2012long} autoencoder to map music features to dance motions. Lee \etal~\cite{lee2019dancing} adopted a different approach, using convolutional neural networks (CNNs) to tackle the problem, while ChoreoMaster~\cite{chen2021choreomaster} introduced motion graphs to provide a more structured and flexible method for dance motion generation. In contrast, speech-to-gesture tasks aim to generate gestures that align with the rhythm and semantics of speech. Ginosar \etal~\cite{ginosar2019learning} used pseudo-labeled data from a motion detection system to train their model, enabling the generation of personalized gesture styles. Aud2Repr2Pose~\cite{kucherenko2019analyzing} developed a joint motion autoencoder and speech encoder, allowing for a more direct mapping between speech features and gesture motion. Audio2Gestures~\cite{li2021audio2gestures} advanced this further by splitting the latent representation into shared and motion-specific components, using carefully designed loss functions to address the one-to-many mapping problem between speech and gestures. Recent works~\cite{ao2023gesturediffuclip,ao2022rhythmic,zhang2024semantic} have leveraged emerging generative models, such as diffusion models~\cite{ho2020denoising} and VQ-VAE~\cite{van2017neural}, to produce rhythm-aware and semantics-aware co-speech gestures and also apply multi-modal data for stylistic variations. Despite these advances, both music and speech signals inherently lack explicit descriptions of the associated human motions~\cite{zhu2023survey}, which limits the interpretability and precision of the generated results. To address this limitation, we propose a novel task centered on generating human motion from explicit audio instructions. By leveraging the richness of verbal descriptions, this approach allows for greater control and precision in the generation process, offering a more intuitive and convenient way of creating human motion.

\textbf{Text-based Motion Generation} focuses on synthesizing human motion sequences that correspond to textual descriptions. Early approaches in this field predominantly used deterministic models to directly map textual inputs to motion outputs. For instance, JL2P~\cite{ahuja2019language2pose} introduced a joint embedding space that aligned motion and text data, ensuring consistency between the two modalities through a reconstruction task. Similarly, Ghosh \etal~\cite{ghosh2021synthesis} employed a Gated Recurrent Unit (GRU)-based model to capture fine-grained motion details, integrating a discriminator to enhance the realism of generated motions. Angela \etal~\cite{lin2018generating} added a trajectory prediction module, improving both complexity and accuracy in motion generation. More recent work has leveraged large pre-trained models to boost performance. MotionCLIP~\cite{tevet2022motionclip}, for example, used the language-image pre-trained model CLIP~\cite{radford2021learning}, tapping into its prior knowledge to improve zero-shot motion generation capabilities. However, these deterministic methods have been limited in producing diverse outputs, as they tend to generate a single plausible motion sequence for each input. To overcome this limitation, stochastic modeling techniques have been introduced to account for the inherently many-to-many nature of the text-to-motion task. TEMOS~\cite{petrovich2022temos}, for instance, replaced traditional deterministic autoencoders with a Variational Autoencoder (VAE), enabling the generation of diverse motion sequences. Text2Action employed a Generative Adversarial Networks (GANs) framework to produce a variety of motions based on textual descriptions. Guo \etal~\cite{guo2022generating} introduced a length predictor alongside a VAE-based generator, allowing the generation of motions with varying lengths that still align with the input descriptions. Building on the success of stochastic models in text-to-image tasks, techniques such as VQ-VAE~\cite{van2017neural} and diffusion models~\cite{ho2020denoising} have been applied to human motion generation. T2M-GPT~\cite{zhang2023generating} used VQ-VAE to learn discrete latent representations of motion sequences, combining this with a GPT-like autoregressive model to predict subsequent motion tokens. Diffusion models, known for their ability to iteratively refine outputs, have also gained attraction. FLAME~\cite{kim2023flame} was the first model to apply diffusion techniques to human motion generation, using free-form language descriptions as input. MDM~\cite{tevet2023human} further advanced this approach by employing a transformer-based diffusion model that predicts the final motion sequence through iterative denoising steps, producing more precise and realistic motions. MoMask~\cite{guo2024momask} further improved efficiency by using a RVQ model for motion representation and masked generative transformers, allowing for faster and more flexible motion generation. However, text-based interaction proves to be inefficient in many practical scenarios, as it demands users to provide exact and comprehensive input, which can be cumbersome and slow. This limitation reduces its suitability for fast-paced applications where intuitive and rapid communication is critical for maintaining a smooth user experience.

% 之前试用的若干种模型
% \textbf{Audio Encoding} has seen significant advancements that transform the landscape of audio processing across various applications. WavLM~\cite{chen2022wavlm} leverages a self-supervised masked speech denoising framework to learn universal speech representations from vast unlabeled datasets, showcasing adaptability across diverse speech processing tasks. Encodec focuses on high-fidelity neural audio compression, employing a residual vector quantization (RVQ)~\cite{borsos2023audiolm} approach to model original audio, enhancing detail recovery in speech and music through a combination of reconstruction loss and discriminator mechanisms. Whisper~\cite{radford2023robust} excels in robust speech recognition by extracting features from the Log-Mel Spectrogram of audio using an Encoder-Decoder framework, achieving impressive multilingual performance with extensive training data. ImageBind~\cite{girdhar2023imagebind} innovatively addresses multimodal alignment through image-paired data, creating a unified representation space for audio, images, and videos, thereby enhancing retrieval and classification tasks in zero-shot and few-shot scenarios. Collectively, these works illustrate the dynamic evolution of audio encoding techniques, enabling higher fidelity, efficiency, and versatility in handling audio data.

\textbf{Large Audio Model} has experienced significant advancements, reshaping audio processing across various applications. WavLM~\cite{chen2022wavlm} employs a self-supervised masked speech denoising framework to learn universal speech representations from extensive unlabeled datasets, demonstrating adaptability across diverse speech processing tasks. Similarly, Encodec~\cite{borsos2023audiolm} utilizes RVQ for high-fidelity neural audio compression, effectively modeling original audio and enhancing detail recovery in both speech and music through a combination of reconstruction loss and discriminator mechanisms. In the realm of speech recognition, Whisper~\cite{radford2023robust} excels by extracting features from the Log-Mel Spectrogram of audio using an encoder-decoder framework, achieving remarkable multilingual performance supported by extensive training data. Furthermore, ImageBind~\cite{girdhar2023imagebind} innovatively addresses multi-modal alignment by leveraging image-paired data to create a unified representation space for audio, images, and videos, thereby enhancing retrieval and classification tasks in zero-shot and few-shot scenarios. Collectively, these developments underscore the dynamic evolution of audio encoding techniques, enabling higher fidelity, efficiency, and versatility in managing audio data.

% \textbf{Generative Models} have been an emerging pattern in recent years, with various approaches emerging to model complex data distributions. Goodfellow \etal~\cite{goodfellow_generative_2020} introduced Generative Adversarial Networks (GANs), which use a generator-discriminator framework to produce realistic data by pitting the two networks against each other. Variational Autoencoders (VAEs)~\cite{kingma2013auto}, proposed by Kingma and Welling, extend traditional autoencoders by assuming a probabilistic latent space, enabling the generation of diverse samples through latent variable sampling. In contrast, flow-based models~\cite{papamakarios2021normalizing} use invertible transformations to map simple distributions to complex data, allowing for exact likelihood estimation and efficient sampling. Diffusion models~\cite{ho2020denoising} generate data by iteratively denoising Gaussian noise into structured outputs, resulting in high-quality results, especially for image generation tasks. Inspired by diffusion processes, masked generative transformers (e.g., MaskGIT~\cite{chang2022maskgit}) adopt an iterative token generation approach, making them more efficient while maintaining gradual generation strategies.

\textbf{Generative Models} have become a central focus in recent years, offering powerful methods to model complex data distributions. Goodfellow \etal~\cite{goodfellow_generative_2020} introduced GANs, which utilize a generator-discriminator framework where the two networks compete, enabling the generation of highly realistic data. In parallel, Kingma and Welling’s Variational Autoencoders (VAEs)~\cite{kingma2013auto} introduced a probabilistic latent space to traditional autoencoders, facilitating the creation of diverse samples by sampling latent variables. Flow-based models\cite{papamakarios2021normalizing}, on the other hand, leverage invertible transformations to map simple distributions into complex data, enabling efficient sampling with exact likelihood estimation. Diffusion models~\cite{ho2020denoising} approach data generation through an iterative process of denoising Gaussian noise, excelling particularly in tasks like image generation due to their high-quality outputs. Building on the success of these models, masked generative transformers, such as MaskGIT~\cite{chang2022maskgit}, have emerged, employing an iterative token generation method that balances efficiency with the gradual refinement of generated content. Collectively, these approaches represent significant strides in generative modeling, each bringing unique strengths to different types of data and tasks.

\section{Our Approach}

\begin{figure*}[h]
    \centering
    \includegraphics[width=\linewidth]{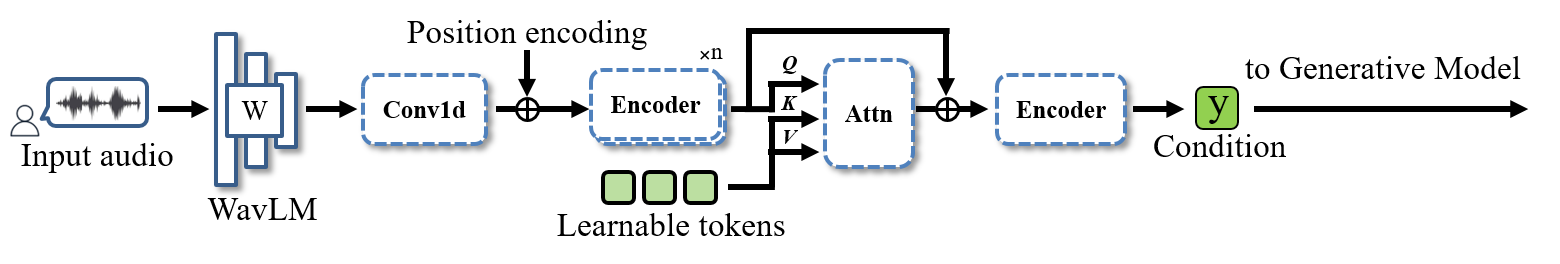}
    \caption{
    \revise{\textbf{Audio Conditions Processing Pipeline.} The audio features extracted by WavLM~\cite{chen2022wavlm} are processed through a memory-retrieval based module, which standardizes the varying lengths of the input audio conditions. This module ensures that all audio signals are converted into a consistent length, facilitating smooth integration with subsequent components in the pipeline. The processed features will be used as conditional signals input into the generative model.}
    }
    % \vspace{-10pt}
    \label{fig:condition_processing_branch}
\end{figure*}

\begin{figure*}[h]
    \centering
    \includegraphics[width=1.0\linewidth]{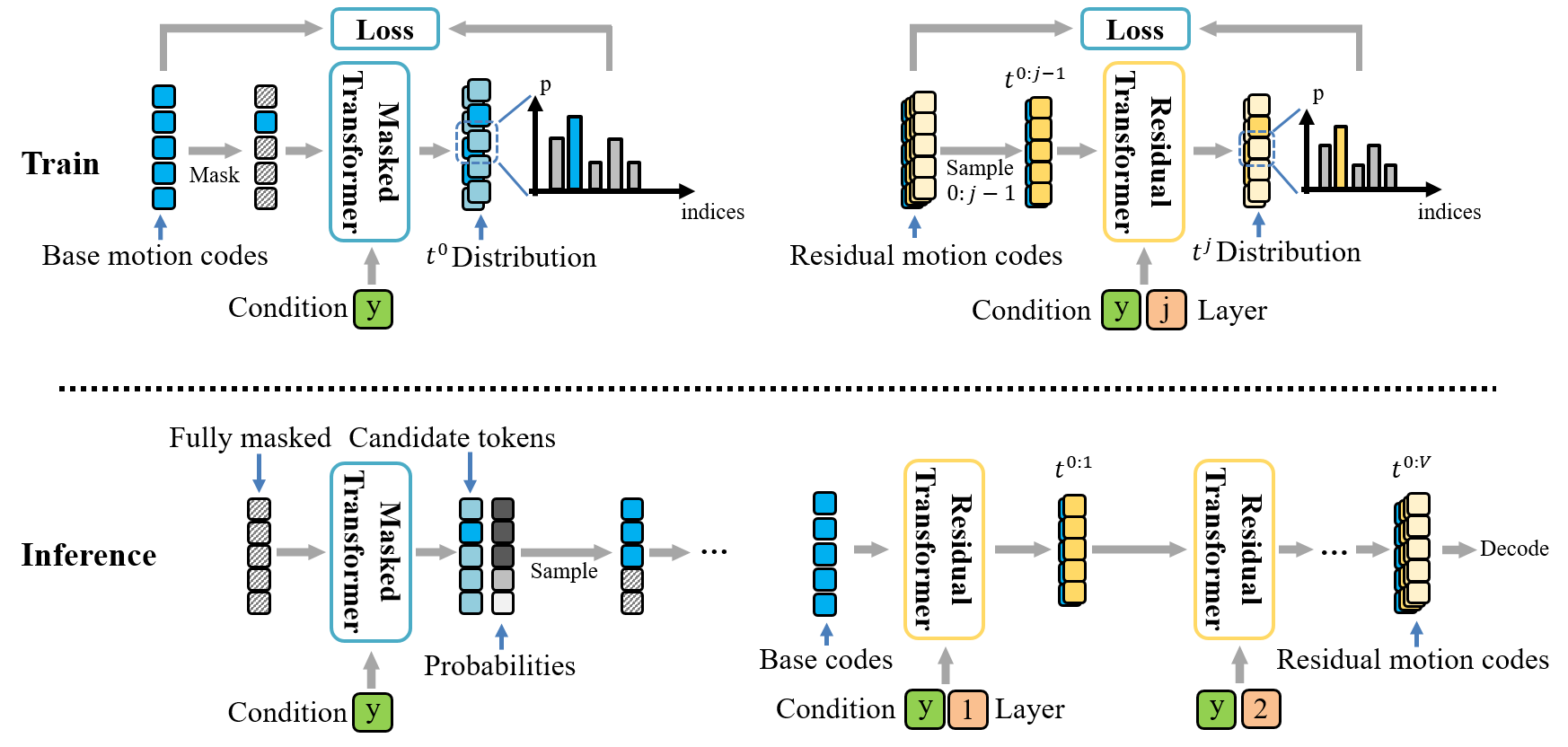}
    \caption{
    \textbf{Overview of Our Generative Framework in Training and Inference.} The framework consists of two key components: The \textbf{Masked Transformer} is designed to model the relationship between audio conditions and the base motion codes, which capture the principal components of the motion. The \textbf{Residual Transformer} establishes the connection between the audio conditions and the residual motion codes, which represent the finer, detailed aspects of the motion. During inference, these transformers work in sequential stages, progressively generating multi-layer latent motion codes that are then decoded to reconstruct the full motion sequence.
    }
    % \vspace{-10pt}
    \label{fig:generative_model_pipeline}
\end{figure*}

In this paper, we propose an end-to-end framework using a masked generative transformer for human motion generation from audio instructions. The process begins by extracting audio features with the pre-trained WavLM model~\cite{chen2022wavlm}. \revise{These features are then compressed into more compact, model-efficient representations using a memory-retrieval based attention module, as depicted in Fig.~\ref{fig:condition_processing_branch}.} Conditioned on these encoded audio features, we train a Masked Transformer to generate the base-level quantization codes for latent motion tokens. This is followed by a Residual Transformer that iteratively generates residual-level quantization codes layer by layer. Finally, the base and residual codes are combined and decoded by the RVQ motion decoder to reconstruct the full motion sequence, as shown in Fig.~\ref{fig:generative_model_pipeline}. This pipeline ensures efficient and accurate motion generation by progressively capturing both high-level structure and detailed nuances.

\subsection{Audio Features Extraction}
\label{subsec:audio_features_extraction}
% \textbf{Audio Signal Processing.} 
% To effectively generate human motion conditioned on audio instructions, a speech vectorization module is employed for vectorization to convert the original audio signals into a data representation suitable for deep learning training. Specifically, this module processes the input sequence of raw audio signals, converting them into vector representations, where the length of the vector sequence corresponds to the input, and the dimensionality of the vectors remains constant. Different audio samples may exhibit varying vector sequence lengths but share the same feature dimensions.

To effectively generate human motion based on audio instructions, we employ a speech vectorization module that converts raw audio signals into suitable feature representations. This module transforms the input audio into vector sequences, ensuring consistent dimensionality for each vector, while allowing the sequence length to vary depending on the input duration. This approach standardizes the feature dimensions across different audio samples, making them compatible with downstream tasks, despite variations in sequence length.

% The vectorization paradigm for the audio modality cannot be directly applied to the text modality due to significant differences in data characteristics between the two modalities, primarily stemming from their inherent properties. Audio signals, as naturally collected data, have lower information density, fewer channels, and longer sequences of sampled vibration points. In contrast, text signals serve as abstract representations with higher information density and clearer semantic features. Moreover, the volume of audio data is typically less than that of text, mainly due to the greater difficulty in data collection and processing, resulting in larger spatial requirements for equivalent semantic content. Therefore, to achieve the fundamental objective of audio-to-motion generation in this study, careful selection of an appropriate audio vectorization scheme is essential for effective feature extraction from the input audio signals, enhancing both the utilization of audio features and semantic extraction capabilities.

\begin{figure}[t]
    \centering
    \includegraphics[width=1.0\linewidth]{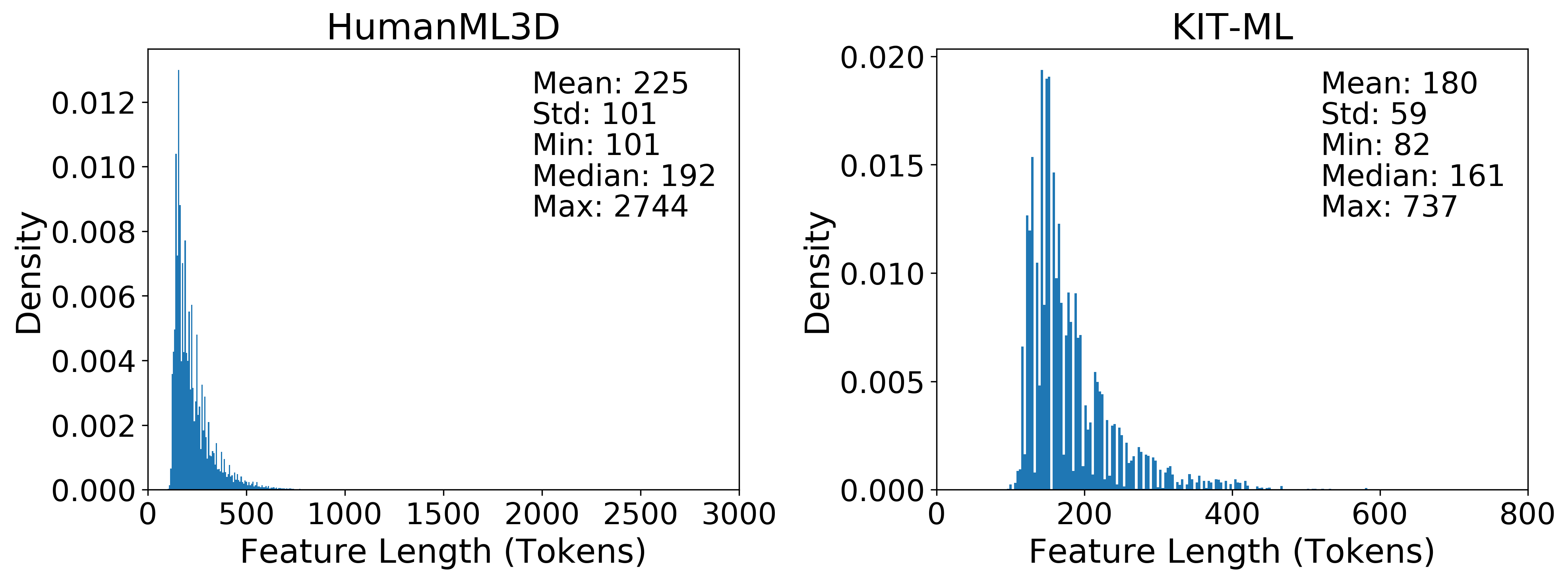}
    \caption{
    \textbf{Distribution of Audio Feature Lengths.} WavLM~\cite{chen2022wavlm} is utilized to extract audio features from the augmented Oral Datasets derived from HumanML3D~\cite{guo2022generating} and KIT~\cite{plappert2016kit}. A statistical analysis of these feature lengths reveals significant variability, with some features exhibiting notably long lengths. This variability presents challenges for processing conditional signals, as it complicates the integration of audio data into subsequent stages of our framework.
    }
    % \vspace{-10pt}
    \label{fig:audio_feature_lengths}
\end{figure}

However, the vectorization process for audio signals cannot be directly applied to text due to significant differences in the inherent characteristics of these modalities. Audio signals, as naturally collected data, typically have lower information density, represent lower-level semantics, and consist of much longer sequences of sampled vibration points. In contrast, text is a more abstract representation with higher information density and a clearer, more explicit semantic structure. Additionally, audio data typically involves smaller volumes compared to text, largely due to the complexity and challenges in collection and processing, which require more spatial resources to convey equivalent semantic content. Therefore, selecting an appropriate audio vectorization scheme is critical for effective feature extraction in audio-to-motion generation, ensuring that the extracted features are semantically rich and fully utilized for downstream motion synthesis.

% To this end, it is crucial to explore previous work on audio processing to uncover basic paradigms and processing experiences in the field. Current research on audio-related tasks can be categorized into three main directions: speech synthesis, speech description, and speech recognition. In speech synthesis, the goal is to generate audio that resembles a speaker’s voice while reciting different text based on a given recording of the speaker reading specific content. For speech description tasks, the objective is to algorithmically describe the content or context of a given audio segment, such as distinguishing between human speech and background sounds like rain. Speech recognition tasks aim to identify the spoken text from an audio recording of a person reading aloud.

To advance our approach, it is important to review prior work in audio processing to uncover key paradigms and insights that can inform our method. Several pre-trained audio encoders, such as Encodec~\cite{borsos2023audiolm}, Whisper~\cite{radford2023robust}, and ImageBind~\cite{girdhar2023imagebind}, have been developed, each tailored to specific tasks like speech synthesis, video retrieval, or speech recognition. However, these task-specific designs often introduce biases into the latent representation of audio signals, limiting their generalizability. In contrast, WavLM~\cite{chen2022wavlm}, a deep learning model developed by Microsoft and pre-trained on over 90,000 hours of human speech data, has demonstrated strong generalization ability across a wide range of downstream tasks. This versatility makes WavLM particularly suitable for the requirements of this study, providing robust and unbiased feature representations for our audio-based tasks.

Consequently, the audio vectorization module in our project is built upon the WavLM pre-trained model. Specifically, the output from the final layer of WavLM is used as the vectorized representation of the audio. For a segment of monaural audio sampled at 16 kHz, the resulting vector sequence length corresponds to the audio duration, while the vector dimensions remain constant. This approach ensures efficient and accurate extraction of audio features necessary for further processing. 

\subsection{Memory-Retrieval Based Audio Encoding}
\label{subsec:memory_retrieval_based}

% The distribution of audio feature lengths across samples in the datasets~\cite{guo2022generating,plappert2016kit} is shown in Fig.~\ref{fig:audio_feature_lengths}. It can be seen that audio features extracted using WavLM exhibit significantly varied sequence lengths, with some sequences being particularly long.  Directly feeding these features into the generative model as conditioning signals could result in excessive computational complexity, and the overly large conditioning signals may also hinder model convergence. To address this issue, it is necessary to encode the audio features into more compact representations that carry richer semantic information. 

The distribution of audio feature lengths across samples in the datasets~\cite{guo2022generating,plappert2016kit} is shown in Fig.~\ref{fig:audio_feature_lengths}. Audio features extracted using WavLM vary significantly in sequence length, with some sequences being notably long. Directly feeding these lengthy features into the generative model as conditioning signals could lead to high computational complexity, while also impeding model convergence due to the overwhelming size of the input. To mitigate these challenges, it is crucial to encode the audio features into more compact representations that not only reduce computational burden but also extract rich motion-related semantic information for effective downstream processing.

\begin{figure*}[htb]
    \centering
    \includegraphics[width=0.85\linewidth]{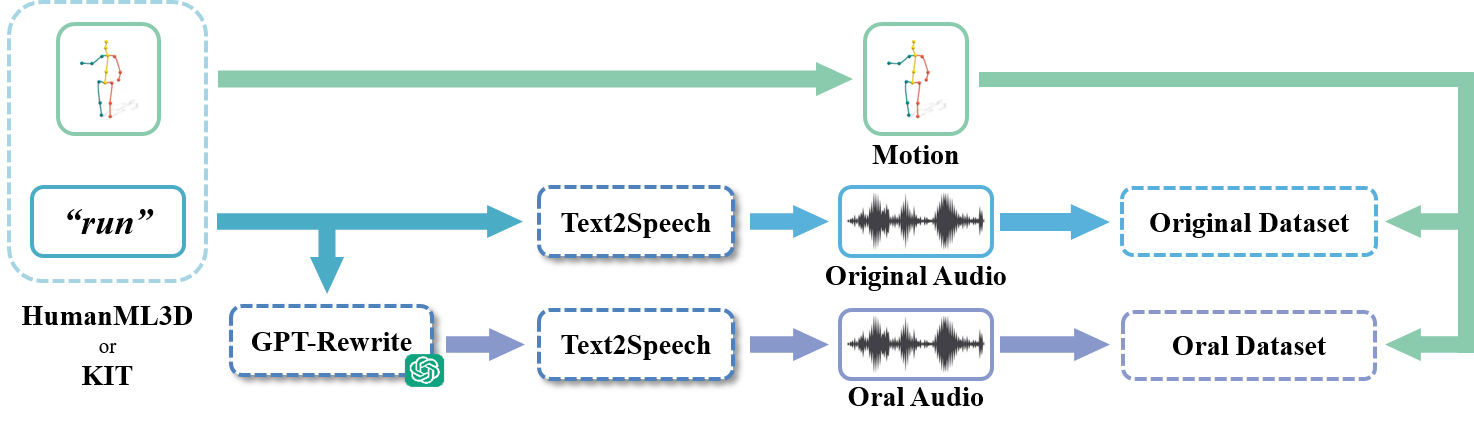}
    \caption{\textbf{Overview of the Audio-Motion Dataset Augmentation Process.} The texts from existing datasets, KIT~\cite{plappert2016kit} and HumanML3D~\cite{guo2022generating}, are fed into the text2speech model, Tortoise~\cite{betker2023better}, generating audio signals with random speaker identities to create the \textbf{Original Dataset}. Additionally, the large language model ChatGPT-3.5~\cite{openai2024gpt} is employed to rewrite the original texts in a more conversational, spoken language style. These rewritten texts are then used to generate corresponding audio signals, forming the \textbf{Oral Dataset}.
    }
    % \vspace{-10pt}
    \label{fig:dataset}
\end{figure*}

A straightforward approach is to adopt the text encoding strategy used by CLIP~\cite{radford2021learning}, which appends a special token to each sequence of audio features and employs a Transformer Encoder~\cite{vaswani2017attention} to aggregate the sequence's features through a self-attention mechanism. Given the significant variation in sequence lengths across different samples, this special token can capture information from any token within the sequence, regardless of its distance, allowing it to effectively model varying-range dependencies. Additionally, a CNN can be applied for downsampling before the features are passed into the Transformer Encoder. This process not only reduces sequence length but also compresses the sparse audio representation into a more compact and semantically enriched form, enhancing both efficiency and feature quality.

% While this approach is more computationally efficient, its performance in our experiments was suboptimal. We believe this is because the tasks of WavLM pre-training are not specifically designed for motion generation, resulting in instructive audio features that still contain information weakly related to motion descriptions, such as speaker identity. This irrelevant information introduces noise and interferes with the semantic extraction of motion-relevant content. To address this issue, inspired by~\cite{jiang2024loopy}, we design a memory retrieval-based module to transform the audio features into forms more closely related to motion control. Specifically, this module stores a set of learnable tokens, which are mapped to the key-value pairs in the attention mechanism. For each input, a fully connected network maps the input to a query vector, and the attention mechanism is used to compute the weighted sum of the learnable tokens as a new feature. After adding positional encodings, this new feature is fed into the next layer of the encoder.

\revise{While this approach offers greater computational efficiency, it yielded suboptimal performance in our experiments. We attribute this to the fact that WavLM's pre-training tasks were not specifically tailored for motion generation, and relying on a single token to represent the rich audio features might create a bottleneck. As a result, the audio features extracted still contain information that is weakly related to motion description, such as speaker identity. This irrelevant information introduces noise, complicating the extraction of motion-relevant semantics. To address this limitation, we draw inspiration from~\cite{jiang2024loopy} and design a memory-retrieval based module that transforms the audio features into representations more closely aligned with motion control. Specifically, this module leverages a set of learnable tokens stored in memory, which are mapped to key-value pairs in the attention mechanism, denoted as $K_{m}$ and $V_{m}$. For each input $x$, a fully connected network generates a query vector $Q_{x}$, and the attention mechanism computes a weighted sum of the learnable tokens to produce a refined feature representation. After positional encodings are added, this refined feature is passed to the next layer of the encoder, allowing for more targeted and relevant semantic extraction that improves motion generation quality. This memory-retrieval process can be described as follows:}
\begin{equation}
\text{Attention}(Q_x, K_m, V_m) = \text{softmax}\left(\frac{Q_xK_m^\top}{\sqrt{d}}\right) V_m,
\label{eq:swap_toy}
\end{equation}
\revise{where $d$ stands for the dimension of the key and value vectors. Through memory-retrieval based audio encoding, we compress variable-length audio features into a compact, fixed-size condition signal, denoted as $y\in\mathbb{R}^{d_y}$, where $d_y$ denotes the dimension of the condition signal. This compact representation serves as the conditioning input for human motion generation, ensuring consistent and efficient use of audio data regardless of its original length.}

\subsection{Masked Generative Model}
\label{subsec:masked_generative_model}
% We follow the practice of Momask~\cite{guo2024momask} to build the generative model. 
% Firstly, we adopt the Residual Vector Quantization (RVQ) method to quantize actions, and then generate actions in a discrete space. 
% Specifically, the motion sequence $m_{1:N}\in\mathbb{R}^{N\times D}$ is firstly encoded into latent sequence $\Tilde{c}_{1:n}\in \mathbb{R}^{n\times d}$, where $n/N$ is the downsampling ratio and $d$ represents the latent dimension. The latent sequence is then represented as an ordered sequence of $V+1$ motion codes $[b^{v}_{1:n}]^V_{v=0}$ through the following process:
% \begin{equation}
%     b^{v}=\operatorname{Q}\left(r^v\right),r^{v+1}=r^v-c^v,v\in [0,V]
% \end{equation}
% Where $r^0=\Tilde{c}$, $\operatorname{Q}\left(\cdot\right)$ stands for traditional quantization operation.
% Through RVQ, the final latent sequence is expressed as the sum of quantized sequences across layers.

% The residual VQ-VAE's loss function encompasses both action reconstruction loss and the loss for each layer's latent space representation:
% \begin{equation}
%     \mathcal{L}_{rvq}=||m-\hat{m}||_1+\beta \sum^{V}_{v=1}||r^{v}-\operatorname{sg[c^v]}||^2_2
% \end{equation}
% Where the $\operatorname{sg[\cdot]}$ denotes the stop-gradient operation, and the $\beta$ a weighting factor for embedding constraint.

\textbf{Latent Motion Representation}  is handled by the RVQ-VAE framework~\cite{borsos2023audiolm,guo2024momask}, which quantizes motion sequences into multi-layer token sequences within a discrete latent space, referred to as motion codes. Specifically, a motion sequence $m_{1:N} \in \mathbb{R}^{N \times D}$ is encoded into a latent representation $\Tilde{c}_{1:n} \in \mathbb{R}^{n \times d}$, where $n/N$ denotes the downsampling ratio and $d$ denotes the latent dimension. Starting from $r^0=\Tilde{c}_{1:n}$, the latent representation is then decomposed into an ordered $V+1$ layers of motion codes $[c^{j}_{1:n}]^V_{j=0}$ through the following process:
\begin{equation}
c^{j}=\operatorname{Q}\left(r^j, C^j\right), \quad r^{j+1}=r^j-c^j, \quad j\in [0, V].
\end{equation}
Here, $\operatorname{Q}\left(\cdot\right)$ represents the standard quantization operation which finds the closet code $c^{j}$ in the codebook $C^j$ as the quantization of the input $r^j$. The codebook $C^j\in\mathbb{R}^{k\times d}$ consists of $k$ codes with each code of dimension $d$. Given the multi-layer motion codes as the latent motion representation, they are summed together as the input to the motion decoder to recover the input motion, denoted as $\hat m_{1:N}$. The overall loss function for the RVQ-VAE consists of two terms: the motion reconstruction loss and the latent embedding loss for each layer:
\begin{equation}
\mathcal{L}_{rvq}=||m-\hat{m}||_1+\beta \sum^{V}_{j=1}\|r^{j}-\operatorname{sg}[c^j]\|^2_2,
\end{equation}
where $\operatorname{sg[\cdot]}$ is the stop-gradient operation and $\beta$ is a weighting factor that regulates the latent codes.

After quantizing the motion sequence using RVQ-VAE, we derive multi-layer motion codes represented as $[c^{j}_{1:n}, t^{j}_{1:n}]^V_{j=0}$, where $t^{j}_{1:n}$ denotes the codebook indices of the quantization code $c^{j}_{1:n}$. Build upon the latent motion representation, our generative model is trained to construct the relationship between audio conditions and base motion code (layer $0$) or residual motion codes (layers $1$ to $V$). The overall framework of the model is illustrated in Fig.~\ref{fig:generative_model_pipeline}, showcasing the training and inference (generation) process.

% \textbf{Masked transformer}, inspired by MaskGIT~\cite{chang2022maskgit}, is designed to model relationship between the audio conditions and the base layer ($v=0$) motion codes $t^0_{1:n}\in \mathbb{R}^n$. During training, a varying fraction of sequence elements are randomly masked out, substituting tokens with a special $\mathrm{[MASK]}$ token. Denoting the masked sequence as $\Tilde{t}^{0}$, the objective is to predict the masked tokens given the condition $y$ and $\Tilde{t^{0}}$. Mathematically, our masked transformer $p(\theta)$ is optimized by minimizing the negative log-likelihood of target predictions:
% \begin{equation}
%     \mathcal{L}_{mask}=\sum_{\Tilde{t}^{0}_k=[MASK]}-\log p_\theta(t^{0}_k|\Tilde{t}^{0},c)
% \end{equation}

\textbf{Masked Transformer}, inspired by MaskGIT~\cite{chang2022maskgit}, is designed to capture the relationship between the audio conditions and the base motion codes $t^0_{1:n}\in \mathbb{R}^n$ ($j=0$). During training, a random fraction of sequence elements is masked, replacing the corresponding tokens with a special $\mathrm{[MASK]}$ token. Letting $\Tilde{t}^{0}$ represent the masked tokens, the model’s task is to predict the masked tokens given the condition $y$ and the partially masked tokens $\Tilde{t^{0}}$. Mathematically, the masked transformer $p_{\theta}$ is trained by minimizing the negative log-likelihood of the target predictions:
\begin{equation}
    \mathcal{L}_{mask}=\sum_{\Tilde{t}^{0}_k=\mathrm{[MASK]}}-\log p_\theta(t^{0}_k|\Tilde{t}^{0}, y).
\end{equation}

\textbf{Residual Transformer} is designed to establish the relationship between the audio conditions and the residual motion codes $t^j_{1:n}\in \mathbb{R}^n$ for $j\in[1,V]$. During training, a quantization layer $j$ is randomly selected. The motion tokens from preceding layers, $t^{0:j-1}$, are collected and summed to create the preceding motion representation as input. Along with the audio conditions $y$ and an indicator embedding for layer $j$, the residual transformer $p_{\phi}$ is trained to predict the tokens for the $j$-th layer. The training objective for the residual transformer is defined as follows:
\begin{equation}
\mathcal{L}_{res}=\sum_{j=1}^{V}\sum_{i=1}^{n}-\log p_\phi(t^{j}_i|t^{1:j-1}_i,y,j)
\end{equation}

\textbf{Motion Generation} is carried out in two stages, as illustrated in Fig.~\ref{fig:generative_model_pipeline}. In the first stage, the goal is to iteratively sample the base motion codes over $L$ iterations. The process begins with fully masked codes, which are fed into the masked transformer. The transformer generates a probability distribution over the base-layer codebook $C^0$, from which candidate motion codes are sampled. At each iteration $l$, the $\left \lceil{\gamma(\frac{l}{L})n}\right \rceil$ tokens with the lowest probabilities are re-masked with the $\mathrm{[MASK]}$ token and regenerated in subsequent iterations. This process continues until all base motion codes are generated by the end of $L$ iterations. The masking ratio $\operatorname{\gamma}(\cdot)$  is dynamically adjusted using a cosine scheduling strategy:
\begin{equation}
    \gamma(x)=\operatorname{cos}\left(\frac{\pi x}{2}\right)\in [0,1]
\end{equation}
Once the base motion codes are generated, the second stage involves progressively generating the residual motion codes using the residual transformer. Over $V$ iterations, the $j$-th iteration takes all previously generated motion codes $t^{1:j-1}$ and the layer number $j$ as inputs to predict the residual motion codes $t^j$ for layer $j$. This process repeats for each subsequent layer, ultimately producing motion codes across all $V+1$ layers. These multi-layer motion codes are then passed into the motion RVQ-VAE decoder, which reconstructs the final generated motion.

During inference, Classifier-Free Guidance (CFG) is applied at the final linear projection layer before the softmax operation. Here, the final logits $w_g$ are computed by adjusting the conditional logits $w_c$ in relation to the unconditional logits $w_u$, using a guidance scale $s$. Specifically, this is done through the formula $w_g=(1+s)\cdot w_c - s\cdot w_u$, which moves the conditional logits further from the unconditional ones.

\section{Dataset Augmentation}
% The dataset augmentation process is illustrated in Fig.~\ref{fig:dataset}. We first utilize the text-to-audio synthesis algorithm to create basic audio-motion (A2M) datasets from traditional text-to-motion datasets (Sec.~\ref{subsec:basedataset}). Subsequently, we generate an oral version with descriptions rewritten through the help of ChatGPT~\cite{openai2024gpt} (Sec.~\ref{subsec:oraldataset}). 

The dataset augmentation process is illustrated in Fig.~\ref{fig:dataset}. First, we apply a text-to-audio synthesis algorithm to transform traditional text-to-motion datasets (see Sec.~\ref{subsec:basedataset}) into what we refer to as the \textbf{Original Dataset}. Next, we create an oral version of the dataset by rewriting the text descriptions into a more conversational style using ChatGPT~\cite{openai2024gpt} (see Sec.~\ref{subsec:oraldataset}), followed by synthesizing corresponding audio files. This results in the formation of the \textbf{Oral Dataset}, which better reflects natural spoken language patterns.

\begin{table}[h]
    \centering
    % \vspace{-10pt}
    \begin{tabularx}{\columnwidth}{|X|X|}
        \hline
        \textbf{Base Datasets} & \textbf{Oral Datasets} \\ 
        \hline
        a person running in circles
        & (1) Run in circles. \\ 
        & (2) Move around in circular motions while running. \\ 
        & (3) Jog in circular patterns. \\ 
        \hline
        a person kicks in the air, runs forward and then trots backward to their original location 
        & (1) I want you to kick in the air, move forward, and then trot back to your starting point. \\
        & (2) Kick in the air, run forward, and trot backward to where you started. \\ 
        & (3) Can you kick in the air, run ahead, and then trot back to your original spot? \\ 
        \hline
        a man squats extraordinarily low then bolts up in an unsatisfactory jump. 
        & (1) Squat down very low and then jump up in an unsatisfactory manner. \\ 
        & (2) Go into an extremely low squat, then jump up in a disappointing leap. \\ 
        & (3) Squat down remarkably low and then spring up in an unsatisfactory jump. \\ 
        \hline
    \end{tabularx}
    % \vspace{3pt}
    \caption{
    \textbf{Examples of Rewritten Descriptions.} The original text descriptions from the Base Datasets were rewritten into a more conversational style, forming the augmented Oral Datasets. This transformation aims to better align the dataset with natural spoken language, enhancing its applicability to real-world voice-based interaction scenarios. 
    }
    \label{tab:gpt_rewritten}
\end{table}

\subsection{Original Dataset}
\label{subsec:basedataset}
% \textbf{Motivation.}
% To achieve human motion generation based on instructive audio signals, a suitable audio-to-motion dataset is essential. The most commonly used datasets for human motion generation under explicit semantic constraints provided by conditional signals are KIT~\cite{plappert2016kit} and HumanML3D~\cite{guo2022generating}. The KIT dataset comprises 3,911 action sequences paired with 6,278 text descriptions. With a larger scale, the HumanML3D dataset contains 14,616 action sequences, representing a diverse range of human behaviors, including dancing and exercising, accompanied by 44,970 descriptive texts. However, despite the abundance of human motion data in these datasets, their conditional signals are textual, making them unsuitable for direct use in instructive audio-conditioned human motion generation tasks. In light of this, we attempt to utilize existing text-to-speech algorithms to synthesize a corresponding audio-motion paired dataset from the paired text-motion dataset, facilitating subsequent work.

For effective human motion generation based on instructive audio signals, an appropriate audio-to-motion dataset is necessary. The most commonly used datasets for motion generation under explicit semantic constraints—such as KIT~\cite{plappert2016kit} and HumanML3D~\cite{guo2022generating}—are text-based. The KIT dataset contains 3,911 action sequences paired with 6,278 text descriptions, while HumanML3D, on a larger scale, includes 14,616 action sequences covering a wide variety of behaviors such as dancing and exercising, paired with 44,970 descriptive texts. Despite the wealth of motion data, both datasets rely on textual conditional signals, which makes them unsuitable for tasks in instructive audio-based motion generation. To address this gap, we leverage existing text-to-speech algorithms to synthesize an audio-motion paired dataset from the original text-motion datasets, providing a foundation for subsequent work in audio-conditioned motion generation.

% \textbf{Basic A2M Datasets Synthesis.} 
% With the rapid advancements in deep learning, speech synthesis technologies have also seen significant progress. Although deep neural network-based speech synthesis algorithms still belong to the realm of traditional statistical parametric synthesis, they leverage the robust fitting capabilities of contemporary neural networks, resulting in natural generation results. Consequently, mainstream text-to-speech synthesis algorithms are predominantly based on deep learning, trained on large-scale text-to-speech datasets, achieving synthesis results that closely approximate real human voices. Among these, Tortoise~\cite{betker2023better}, which is grounded in a denoising diffusion probabilistic model~\cite{ho2020denoising} that has gained widespread success in the field of image generation, sacrifices some operational efficiency through an iterative generation process to produce more realistic synthesis outcomes. 

Recent speech synthesis technologies have also seen significant progress. While deep learning based speech synthesis still falls under the category of traditional statistical parametric synthesis, the powerful fitting capabilities of modern neural networks enable highly natural speech generation. As a result, most mainstream text-to-speech synthesis algorithms are now deep learning-based, trained on large-scale text-to-speech datasets to achieve synthesis that closely mirrors real human voices. Among these models, Tortoise~\cite{betker2023better} stands out. Based on a denoising diffusion probabilistic model~\cite{ho2020denoising}, which has found significant success in image generation, Tortoise sacrifices some efficiency due to its iterative process in favor of producing highly realistic speech.

% Specifically, for a given text input, the audio synthesized by Tortoise not only resembles human voices in terms of timbre and quality but also incorporates details such as natural pauses, emphasis, and tonal variations based on the semantic nuances of the text, thus enhancing the naturalness of the generated speech to meet practical application requirements. Furthermore, Tortoise features a probabilistic generation framework allowing for the production of voices from various speakers, yielding richer outputs that cater to a wider array of application scenarios. During the synthesis process using Tortoise, since it does not support the explicit specification of speakers, we generate a diverse set of data featuring randomly selected speakers. Additionally, to facilitate the subsequent algorithm's audio processing, the synthesized audio is stored in mono WAV format with a fixed sampling rate of 16 kHz. To expedite the synthesis process, GPUs are employed for acceleration. The data synthesis process is illustrated in the accompanying figure.

For a given text input, Tortoise synthesizes audio that not only matches human voices in timbre and quality but also captures natural features like pauses, emphasis, and tonal variation, reflecting the semantic intricacies of the text. This attention to detail enhances the naturalness of the generated speech, making it suitable for practical applications. Additionally, Tortoise’s probabilistic generation framework allows for the synthesis of voices from multiple speakers, enriching its outputs and making it adaptable to diverse scenarios. Since Tortoise does not support specifying speakers directly, we generate data using randomly selected voices, adding variability to the dataset.

To support downstream audio processing algorithms, all synthesized audio is saved in mono WAV format with a fixed sampling rate of 16 kHz. The synthesis process is accelerated using GPUs to expedite the generation of large-scale data. The steps involved in this data synthesis process are illustrated in the accompanying figure.

\subsection{Oral Dataset}
\label{subsec:oraldataset}
% \textbf{Oral A2M Datasets Synthesis.} 
% Considering the practical usage scenarios of this task, users' wording and syntactic habits during voice input may differ significantly from those during text input. To align the dataset more closely with real-world applications, this study attempts to use a large language model to modify the descriptive texts for a more conversational tone, thereby narrowing the gap between the dataset's descriptive texts and potential spoken language descriptions. Among the large language models, ChatGPT~\cite{openai2024gpt} presented by OpenAI has demonstrated excellent results in English natural language understanding and instruction following. Therefore, we employ ChatGPT 3.5 to rewrite the descriptions. Specifically, by designing prompts that constrain the format of the generated text, we require that the content remains aligned with the original description while showcasing more diverse sentence structures and vocabulary that align with the conversational habits of users. Table~\ref{tab:gpt_rewritten} exhibits some motion descriptions rewritten utilizing ChatGPT. The rephrased descriptions exhibit greater syntactic and lexical diversity while preserving congruent content with the original descriptions, based on which Tortoise is applied to generate corresponding audio. Ultimately, 12696 oral audio-motion data pairs are synthesized for the KIT-ML dataset and 87384 for HumanML3D.

Considering the practical usage scenarios for this task, users' phrasing and syntactic patterns during voice input often differ significantly from those used in text input. To better align the dataset with real-world applications, this study aims to modify the descriptive texts to reflect a more conversational tone, thereby reducing the gap between the dataset's written descriptions and potential spoken language input. Among the large language models available, OpenAI’s ChatGPT~\cite{openai2024gpt} has demonstrated exceptional performance in English natural language understanding and instruction following. Therefore, we utilize ChatGPT 3.5 to rewrite the motion descriptions.

By crafting prompts that guide the model to maintain alignment with the original content, we ensure that the generated text retains the core meaning while incorporating more varied sentence structures and vocabulary, better reflecting conversational habits. Tab.~\ref{tab:gpt_rewritten} displays some examples of motion descriptions rewritten using ChatGPT, showing greater syntactic and lexical diversity without altering the original meaning. Based on these revised descriptions, the Tortoise model is then applied to generate the corresponding audio. Ultimately, this process yields 12,696 oral audio-motion pairs for the KIT-ML dataset and 87,384 for HumanML3D.

\section{Experiment}
\label{sec:experiment}
% Empirical assessments will be carried out using our speech-motion paired dataset synthesized from HumanML3D~\cite{guo2022generating} and KIT~\cite{plappert2016kit}. For both datasets, pose representations derived from T2M~\cite{guo2022generating} are adopted.

\begin{table*}[h]
    % \vspace*{-10pt}
    \centering
    \renewcommand{\arraystretch}{1.0}
    \resizebox{\textwidth}{!}{
    \begin{tabular}{l c c c c c c c c}
    \toprule
    \multirow{2}{*}{\textbf{Datasets}} & \multirow{2}{*}{\textbf{Method}} & \multirow{2}{*}{\makecell[c]{\textbf{Training}\\\textbf{Data}}}  & \multicolumn{3}{c}{\textbf{R Precision}$\uparrow$} & \multirow{2}{*}{\textbf{FID}$\downarrow$} & \multirow{2}{*}{\textbf{MM Dist}$\downarrow$} & \multirow{2}{*}{\textbf{MultiModality}$\uparrow$}\\

    \cline{4-6}
       ~ & ~ & ~ & Top 1 & Top 2 & Top 3 \\
    \midrule
    \multirow{4}{*}{\makecell[c]{Human\\ML3D}} 

        ~& \multirow{2}{*}{\makecell[c]{Momask~\cite{guo2024momask}\\(Text-Based)}} & Original& \et{0.400}{.002} & \et{0.583}{.003} & \et{0.689}{.002} & \et{0.251}{.009} & \et{3.808}{.009} & \etb{1.407}{.049}  \\ 
        
        ~& & Oral & \etb{0.431}{.002} & \etb{0.626}{.002} & \etb{0.735}{.002} & \etb{0.091}{.003} & \etb{3.551}{.006} & \et{1.284}{.042}  \\ 
    \cline{2-9}
        ~& \multirow{2}{*}{\makecell[c]{Ours\\(Audio-Based)}} & Original& \et{0.417}{.004} & \et{0.603}{.003} & \et{0.708}{.002} & \et{0.202}{.011} & \et{3.697}{.018} & \etb{1.272}{.041}  \\ 
        
    % \cline{2-8}
        ~& & Oral & \etb{0.426}{.007} & \etb{0.619}{.009} & \etb{0.727}{.009} & \etb{0.126}{.005} & \etb{3.577}{.037} & \et{1.222}{.036}  \\
        
    % \bottomrule
    \midrule
    \multirow{4}{*}{\makecell[c]{KIT-\\ML}} 

        ~& \multirow{2}{*}{\makecell[c]{Momask~\cite{guo2024momask}\\(Text-Based)}} & Original& \et{0.309}{.008} & \et{0.513}{.009} & \et{0.652}{.009} & \etb{0.258}{.017} & \et{4.017}{.064} & \etb{1.562}{.064}  \\ 
        
        ~& & Oral & \etb{0.330}{.006} & \etb{0.536}{.007} & \etb{0.667}{.008} & \et{0.261}{.014} & \etb{3.840}{.033} & \et{1.166}{.037}  \\ 
    \cline{2-9}
        ~& \multirow{2}{*}{\makecell[c]{Ours\\(Audio-Based)}} & Original& \et{0.314}{.008} & \et{0.508}{.012} & \et{0.636}{.017} & \et{0.487}{.046} & \et{4.186}{.140} & \etb{1.399}{.052}  \\ 
    % \cline{2-8}
        % ~& \textbf{Oral} & \et{0.324}{.005} & \et{0.529}{.006} & \et{0.667}{.006} & \et{0.207}{.010} & \et{3.899}{.027} & \et{1.312}{.048}  \\
        % momask original code
        ~& & Oral & \etb{0.324}{.005} & \etb{0.529}{.006} & \etb{0.667}{.006} & \etb{0.207}{.010} & \etb{3.899}{.027} & \et{1.312}{.048}  \\
    \bottomrule
    \end{tabular}
    }
    % \vspace{3pt}
    \caption{
    \textbf{Quantitative Evaluation on Oral Datasets.} Both the text-based method, Momask, and our audio-based method are trained on either the original or oral datasets and tested on the oral datasets. The results underscore the importance of our oral-style datasets as training data in better aligning with conversational applications. Furthermore, the evaluation highlights the effectiveness of audio signals for semantic conditioning, showing similar performance to text-based conditions. The notation $\pm$ indicates a 95\% confidence interval.
    }
    \label{tab:quantative_oral_dataset}

\end{table*}

\begin{figure*}[h]
    \centering
    \includegraphics[width=0.7\textwidth]{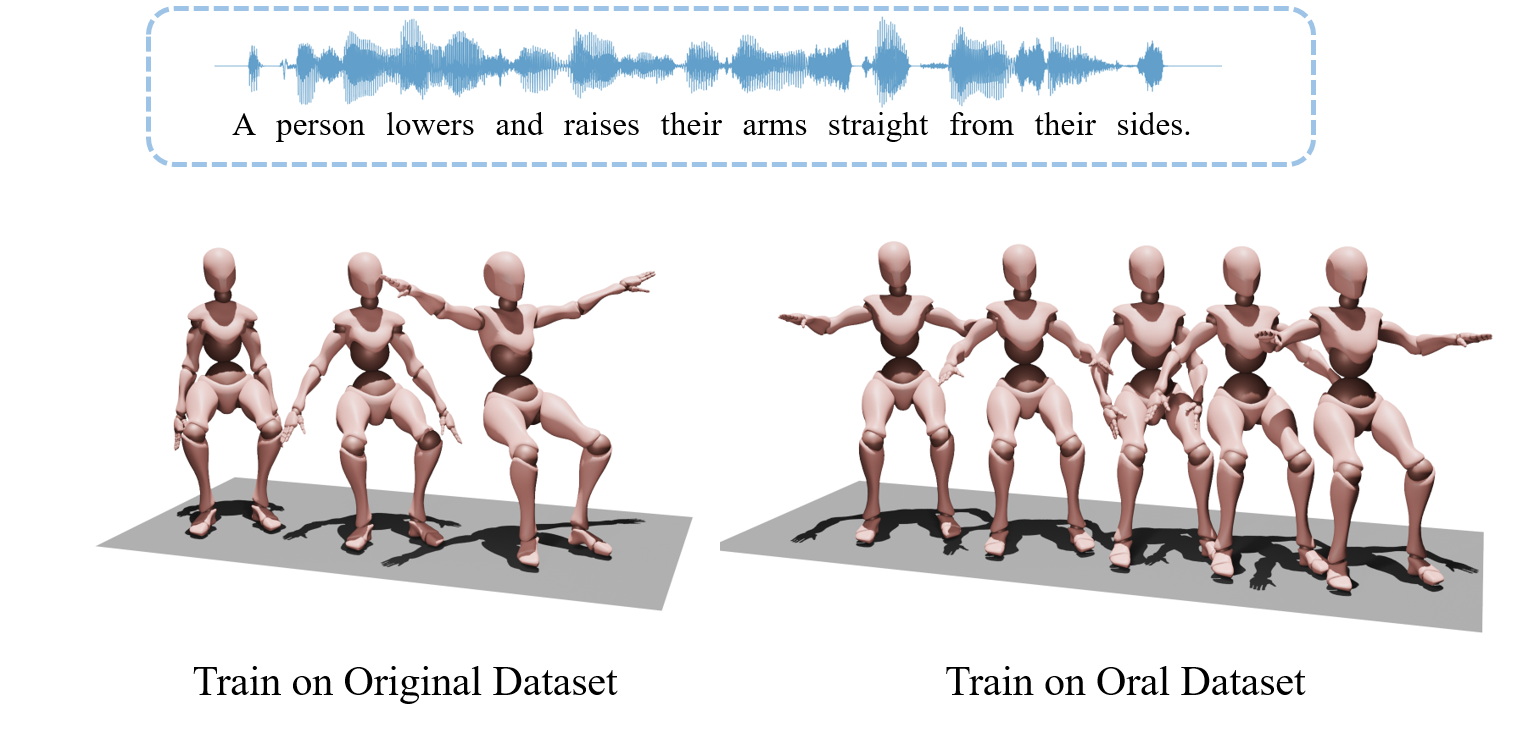}
    \caption{
    \textbf{Qualitative Comparison between Models Trained on Original and Oral Datasets}. As illustrated in the figure, a comparison of the performance of models trained on the Oral and Original datasets reveals that the model trained on the Oral dataset demonstrates greater robustness to audio instructions. Which can be attributed to the higher diversity of the training data.
    }
    % \vspace{-10pt}
    \label{fig:original-oral_performance}
\end{figure*}

Empirical evaluations will be conducted using our augmented Oral Datasets and Base Datasets from HumanML3D~\cite{guo2022generating} and KIT~\cite{plappert2016kit}, with pose representations adopted from T2M~\cite{guo2022generating}. These datasets serve as the foundation for assessing the performance of our model in generating motion sequences from audio instructions.

In our study, we adopt a set of evaluation metrics derived from both T2M~\cite{guo2022generating} and MotionDiffuse~\cite{zhang2022motiondiffuse} to ensure a comprehensive quantitative assessment of our results. These metrics collectively provide insights into the generation quality, semantic coherence, and diversity of our output, including:
\begin{itemize}
    \item \textbf{Frechet Inception Distance (FID)}: Measures the quality of generated motions by comparing the distributional differences between the high-level features of the generated and real motion sequences.
    \item \textbf{R-precision}: Evaluates the semantic consistency between the input conditions (audio/text) and the generated motion.
    \item \textbf{Diversity}: Assesses the variability and richness of the generated action sequences, ensuring the model does not produce overly similar outputs.
    \item \textbf{Multi-modal Distance (MM Dist)}: Computes the average Euclidean distance between the motion features and their corresponding conditioning features, capturing how well the generated motion aligns with the input conditions.
    \item \textbf{MultiModality}: Evaluates the range of distinct motion sequences generated from a single input condition, emphasizing the model's ability to generate multiple plausible outputs for the same input.
\end{itemize}
% (1) Frechet Inception Distance (FID) quantifies the quality of generated motions by assessing the distributional differences between the high-level features of generated and real motion sequences; 
% (2) R-precision, which evaluates the semantic alignment between the condition and the generated motion;
% (3) Diversity, which measures the variability and richness of the generated action sequences; 
% (4) Multimodality, which assesses the average variance of motion sequences generated from a given condition input; and 
% (5) Multi-modal Distance (MM Dist), which calculates the average Euclidean distance between motion features and their corresponding text description features. 

% \textcolor{red}{Implementation Details.}
% \textbf{Implementation Details.} The number of learnable tokens in our memory-retrieval based audio encoding is 256, with the dimension of 512. During training, we set the drop rate of the audio condition to 0.2 and adopt a linear warm-up schedule, converging to a learning rate of 2e-4 after 2000 iterations. Both the masked transformer and residual transformer are composed of six transformer layers, each with six attention heads and a latent dimension of 384, and are applied to the HumanML3D and KIT-ML datasets. During generation process, we set the classifier-free guidance scale to 4 and 5 for the masked and residual transformer on HumanML3D, and (2, 5) on KIT-ML, with a fixed sequence length of 10 for both datasets.

\revise{\textbf{Implementation Details.} The number of learnable tokens in our memory-retrieval based audio encoding is 256, with the dimension of 512. During training, we set the drop rate of the audio condition to 0.2 and adopt a linear warm-up schedule, converging to a learning rate of 2e-4 after 2000 iterations. Both the masked transformer and residual transformer are composed of six transformer layers, each with six attention heads and a latent dimension of 384, and are applied to the HumanML3D and KIT-ML datasets. During generation process, we set the classifier-free guidance scale to 4 and 5 for the masked and residual transformer on HumanML3D, and (2, 5) on KIT-ML, with a fixed sequence length of 10 for both datasets. For visualization, we render the motions using the free character library models provided by Mixamo.}

\begin{table*}[h]
    % \vspace*{-10pt}
    \centering
    \renewcommand{\arraystretch}{1.1}
    % \scalebox{0.9}{
    \resizebox{0.95\linewidth}{!}{
    \begin{tabular}{l l c c c c c c}
    \toprule
    % \multirow{2}{*}{Methods} & \multicolumn{4}{c}{\DN (Coarse-grained)} & & \multicolumn{4}{c}{HumanAct(Fine-grained)} \\
    % \cline{2-5}
    % \cline{7-10}
    %                 & FID$\downarrow$ & Accuracy$\uparrow$ & Diversity$\rightarrow$& MModality$\rightarrow$ &  & FID$\downarrow$ & Accuracy$\uparrow$ & Diversity$\rightarrow$ & MModality$\rightarrow$\\
    \multirow{2}{*}{\textbf{Datasets}} & \multirow{2}{*}{\textbf{Methods}}  & \multicolumn{3}{c}{\textbf{R Precision}$\uparrow$} & \multirow{2}{*}{\textbf{FID}$\downarrow$} & \multirow{2}{*}{\textbf{MM Dist}$\downarrow$} & \multirow{2}{*}{\textbf{MultiModality}$\uparrow$}\\

    \cline{3-5}
       ~& ~ & Top 1 & Top 2 & Top 3 \\
    %\midrule
     %   \textbf{Real motions} & \et{0.511}{.003} & \et{0.703}{.003} & \et{0.797}{.002} & \et{0.002}{.000} & \et{2.974}{.008} & \et{9.503}{.065} & -  \\
    \midrule
    \multirow{9}{*}{\makecell[c]{Human\\ML3D}} & TM2T~\cite{guo2022tm2t} & \et{0.424}{.003} & \et{0.618}{.003} & \et{0.729}{.002} & \et{1.501}{.017} & \et{3.467}{.011} & \ets{2.424}{.093}  \\ 
        ~& T2M~\cite{guo2022generating} & \et{0.455}{.003} & \et{0.636}{.003} & \et{0.736}{.002} & \et{1.087}{.021} & \et{3.347}{.008} & \et{2.219}{.074}  \\   
        ~& MDM~\cite{tevet2023human} & - & - & \et{0.611}{.007} & \et{0.544}{.044} & \et{5.566}{.027} & \etb{2.799}{.072}  \\

        ~ & MLD~\cite{chen2023executing} & \et{0.481}{.003} & \et{0.673}{.003} & \et{0.772}{.002} & \et{0.473}{.013} & \et{3.196}{.010} & \et{2.413}{.079}  \\
        ~ & MotionDiffuse~\cite{zhang2022motiondiffuse} & \et{0.491}{.001} & \et{0.681}{.001} & \et{0.782}{.001} & \et{0.630}{.001} & \et{3.113}{.001}  & \et{1.553}{.042}  \\

        ~ & T2M-GPT~\cite{zhang2023generating} & \et{0.492}{.003} & \et{0.679}{.002} & \et{0.775}{.002} & \et{0.141}{.005} & \et{3.121}{.009}  & \et{1.831}{.048}  \\

        ~ & ReMoDiffuse~\cite{zhang2023remodiffuse} & \et{0.510}{.005} & \ets{0.698}{.006} & \et{0.795}{.004} & \ets{0.103}{.004} & \et{2.974}{.016} & \et{1.795}{.043}  \\
        
        ~ & Momask~\cite{guo2024momask} & \etb{0.521}{.002} & \etb{0.713}{.002} & \ets{0.807}{.002} & \etb{0.045}{.002} & \ets{2.958}{.008} & \et{1.241}{.040} \\
    \cline{2-8}
        % ~ & \textbf{Ours*} & \etb{0.509}{.003} & \etb{0.701}{.002} & \etb{0.795}{.002} & \etb{0.064}{.002} & \etb{3.014}{.018} & \et{1.167}{.030}  \\
        % transformer mean:
        ~ & \textbf{Ours*} & \ets{0.519}{.004} & \etb{0.713}{.005} & \etb{0.808}{.005} & \et{0.121}{.004} & \etb{2.955}{.011} & \et{1.221}{.032}  \\
        
    % \bottomrule
    \midrule
    \multirow{9}{*}{\makecell[c]{KIT-\\ML}} & TM2T~\cite{guo2022tm2t} & \et{0.280}{.005} & \et{0.463}{.006} & \et{0.587}{.005} & \et{3.599}{.153} & \et{4.591}{.026} & \etb{3.292}{.081}  \\ 
        ~& T2M~\cite{guo2022generating} & \et{0.361}{.005} & \et{0.559}{.007} & \et{0.681}{.007} & \et{3.022}{.107} & \et{3.488}{028} & \et{2.052}{.107}  \\   
        ~& MDM~\cite{tevet2023human} & - & - & \et{0.396}{.004} & \et{0.497}{.021} & \et{9.191}{.022} & \et{1.907}{.214}  \\

        ~& MLD~\cite{chen2023executing} & \et{0.390}{.008} & \et{0.609}{.008} & \et{0.734}{.007} & \et{0.404}{.027} & \et{3.204}{.027} & \ets{2.192}{.071}  \\
        
        ~& MotionDiffuse~\cite{zhang2022motiondiffuse} & \et{0.417}{.004} & \et{0.621}{.004} & \et{0.739}{.004} & \et{1.954}{.062} & \et{2.958}{.005}  & \et{0.730}{.013}  \\

        ~& T2M-GPT~\cite{zhang2023generating} & \et{0.416}{.006} & \et{0.627}{.006} & \et{0.745}{.006} & \et{0.514}{.029} & \et{3.007}{.023}  & \et{1.570}{.039}  \\

        ~& ReMoDiffuse~\cite{zhang2023remodiffuse} & \ets{0.427}{.014} & \et{0.641}{.004} & \et{0.765}{.055} & \ets{0.155}{.006} & \ets{2.814}{.012} & \et{1.239}{.028}  \\
        
        ~& Momask~\cite{guo2024momask} & \etb{0.433}{.007} & \etb{0.656}{.005} & \etb{0.781}{.005} & \et{0.204}{.011} & \etb{2.779}{.022} & \et{1.131}{.043}  \\
    \cline{2-8}
        ~& \textbf{Ours*} & \et{0.426}{.006} & \ets{0.645}{.007} & \ets{0.771}{.005} & \etb{0.113}{.004} & \et{2.817}{.021} & \et{1.152}{.048}  \\
    \bottomrule
    \end{tabular}
    }
    % \vspace{3pt}
    \caption{\textbf{Quantitative Evaluation on the Original Datasets.} Previous baseline models are conditioned on text descriptions from the HumanML3D and KIT-ML test sets. In contrast, our method (denoted as Ours\textbf{*}) is conditioned on audio from the Original Datasets. The symbol $\pm$ represents the 95\% confidence interval. Results highlighted in \textbf{bold} indicate the best performance, while those \underline{underlined} represent the second-best. These results underscore the effectiveness of using audio signals as a conditioning input, showing that audio can be as applicable as text in human motion generation tasks.
    }
    \label{tab:quantative_compare_t2m}
\end{table*}

\subsection{Results on Oral Dataset}
\label{subsec:oral_dataset_results}
% 1. distribution difference between original style and oral style.
% 2. audio-based method achieve similar performance with text-based method

% As previously discussed (Section~\ref{subsec:oraldataset}), to better emulate the language patterns employed by users in voice-based interactions, we employed a rewriting approach grounded in ChatGPT~\cite{openai2024gpt} to convert the original dataset into a colloquial form. Subsequently, we utilized Tortoise to transcribe this textual data into corresponding speech. On this synthesized oral dataset, we conducted tests with the proposed end-to-end action generation model. 

As discussed in Sec.~\ref{subsec:oraldataset}, texts in the original datasets often differ significantly from the phrasing and syntactic patterns users employ during audio input. To highlight this, we present the performance of the most recent text-based model, Momask\cite{guo2024momask}, and our audio-based method, both trained on either the original or oral datasets, and tested on the oral datasets. In terms of quantitative results, following previous conventions~\cite{guo2022generating,yuan2023physdiff}, each experiment is repeated 20 times, and the reported metric values represent the mean along with a 95\% statistical confidence interval. 

% And the qualitative results are shown in Fig.~\ref{fig:audio_performance}.
The quantitative results, shown in Tab.~\ref{tab:quantative_oral_dataset}, reveal two key findings: (1) Models trained on the original datasets—whether conditioned on text or audio—perform significantly worse when tested on oral datasets whose the semantics representation is more conversational. This underscores the need for our proposed oral datasets to better align with practical applications. As shown in Fig.~\ref{fig:original-oral_performance}, it is evident that for some specific audio instructions, the models trained on the Oral dataset yield more instruction-congruent generation outcomes. We attribute this to the higher diversity inherent in the Oral dataset, which enhances the robustness of the trained models to handle various forms of instructions. (2) Compared to the text-based method, our audio-based approach presents similar performance regardless of being trained on the original or oral datasets, supporting our hypothesis that audio signals can serve as an effective alternative to text for semantic conditioning. Moreover, audio signals offer more convenient and efficient interactions in practical applications. 

\revise{Notably, as observed in Tab.~\ref{tab:quantative_oral_dataset}, our proposed algorithm yields comparable outcomes to Momask~\cite{guo2024momask} across various metrics, albeit with a less favorable performance in terms of Multimodality. However, it is crucial to highlight that our algorithm leverages audio instructions synthesized from text as input, in contrast to Momask, which benefits from ground truth text input. Given the inherent noise introduced by speech synthesis, it is justifiable that the results in Table 2 do not exhibit a substantial enhancement relative to Momask.}

% \revise{To evaluate the robustness of the proposed model to the audio quality of input speech instructions in real-world application scenarios, this paper uses real recorded speech instructions and adds noise to simulate speech instructions in in-the-wild settings for testing the model. The results are shown in the figure. As seen in Fig.~\ref{fig:in_the_wild}, the model is able to correctly generate actions that align with the command requirements, regardless of whether the audio is synthetic, from real users, or contains noise, demonstrating the robustness of the proposed method.}

\begin{figure*}[h]
    \centering
    \includegraphics[width=0.9\textwidth]{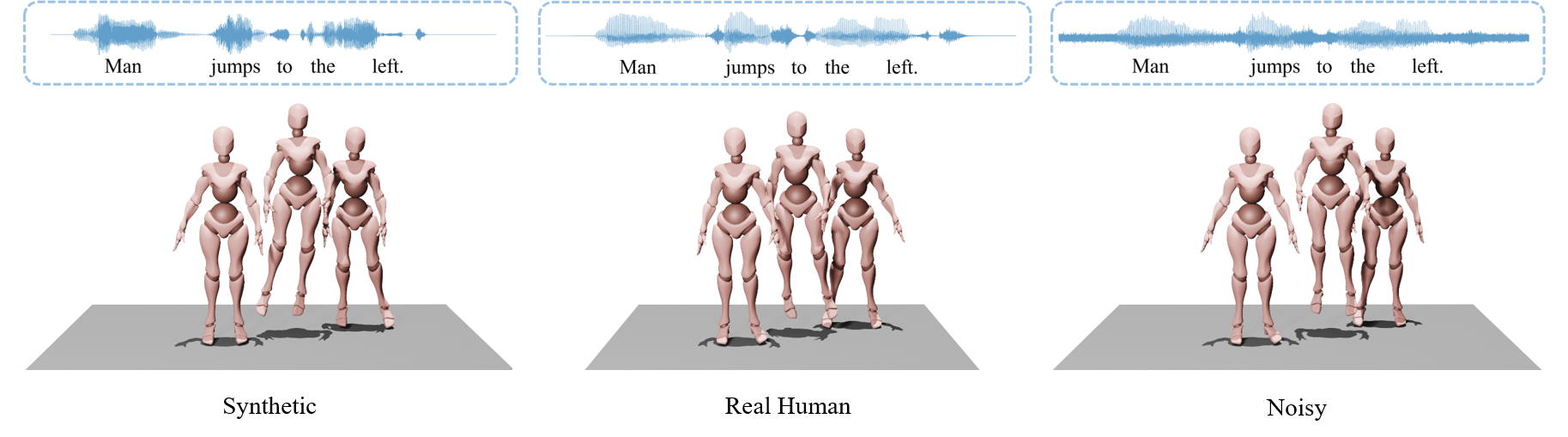}
    
    \caption{\revise{\textbf{Qualitative result with different audio instructions.} Under conditions of synthetic audio, real user  input, and noisy speech instruction, the proposed method is able to correctly generate actions that align with the command requirements, demonstrating the robustness of the proposed method.
    }
    }
    % \vspace{-10pt}
    \label{fig:in_the_wild}
\end{figure*}

\reviseminor{To evaluate the robustness of the proposed model to variations in speech quality under real-world conditions, we conducted additional experiments using real recorded audio instructions. Specifically, the audio samples were recorded in a quiet indoor environment using a microphone at a sampling rate of 12 kHz. To simulate in-the-wild scenarios, white Gaussian noise was added to the recordings with a signal-to-noise ratio (SNR) of 10 dB. A total of three audio instructions were tested under three conditions: synthetic, real recorded, and noisy inputs. Fig.~\ref{fig:in_the_wild} presents one representative example. In general, the model successfully generates motion sequences that align with the intended semantics of the instructions, regardless of the input condition. These results demonstrate the robustness of the proposed method across diverse and practical application settings.
}

% To validate whether a model trained on the oral dataset better aligns with real-world user scenarios, we compared it with a model trained on the base version of the dataset. The results are depicted in Tab.~\ref{tab:quantative_oral_dataset}. From the table, it is evident that the model trained on the oral dataset exhibits superior performance across multiple metrics, demonstrating its enhanced applicability in spoken language contexts.

\subsection{Comparison to Text-Based Methods}
\label{sec:comparision_text_to_motion}
% To further validate the potential of instructive audio in providing explicit motion descriptions comparable to text, this study compares the performance of the proposed algorithm on the synthesized dataset with the popular Text2Motion algorithm. It is important to highlight that the Text2Motion algorithm is trained and evaluated on a text-motion dataset, whereas our Audio2Motion model is developed and assessed using our synthesized audio-motion dataset.

% To further validate the applicability of audio signals in human motion generation, we compare our method with a few text-based methods. The experiments are conducted with original datasets, where our method takes the audio signals as input condition and compared methods use text descriptions. Quantitative results are presented in Table~\ref{tab:quantative_compare_t2m}. Overall, our Audio2Motion model proposed in this paper demonstrates competitive performance in metrics such as FID, R-Precision, and multimodal distance when compared to the most recent methods that utilize text as a condition. These results suggest that instructive audio possesses a representational capability similar to that of text, enabling it to provide explicit descriptions of human movements.

To further assess the effectiveness of audio signals in human motion generation, we compare our approach with several state-of-the-art text-based methods. The experiments are conducted using the original datasets, where our method utilizes audio signals as the input condition, while the compared methods are conditioned on text descriptions. Quantitative results are shown in Tab.~\ref{tab:quantative_compare_t2m}. Our model achieves competitive performance across key metrics, including FID, R-Precision, and multimodal distance, when compared to the latest text-based approaches. 
\reviseminor{
Considering that text-based methods take direct precise textual input, audio instruction-based methods are inherently more challenging.
These results indicate that instructive audio can serve as an equally effective condition as text, offering a comparable representational capacity for describing human movements explicitly, and also providing more convenient user interactions.
}

\subsection{Ablation Study}

In this section, we undertake an ablation study to examine the influence of the key components of our model on its overall performance.

% \subsection{Comparison to Cascaded Method}
% \label{subsec:comparision_to_baseline}
% \textbf{Comparison to Cascaded Method.} The above experiments have demonstrated that audio signals have equivalent semantic representation ability with text in the task of motion generation. Considering that many real-world applications employ cascaded approach that first converts audio to text and then forwards to the downstream tasks. However, if the semantics encoding of audio signals is effective enough, our proposed end-to-end framework will be much more efficient than the cascaded method. Specifically, we employed the small version of Whisper for speech recognition, with the resulting output texts directly fed into our text-based framework for inference, where texts are encoded with CLIP. The performance testing was conducted on the same NVIDIA A6000 hardware. The results are presented in Tab.~\ref{tab:comparison_efficiency}, where our end-to-end generation framework has over a 50\% improvement in generation efficiency compared to the two-stage cascaded method.

\begin{figure*}[h]
    \centering
    \includegraphics[width=0.7\textwidth]{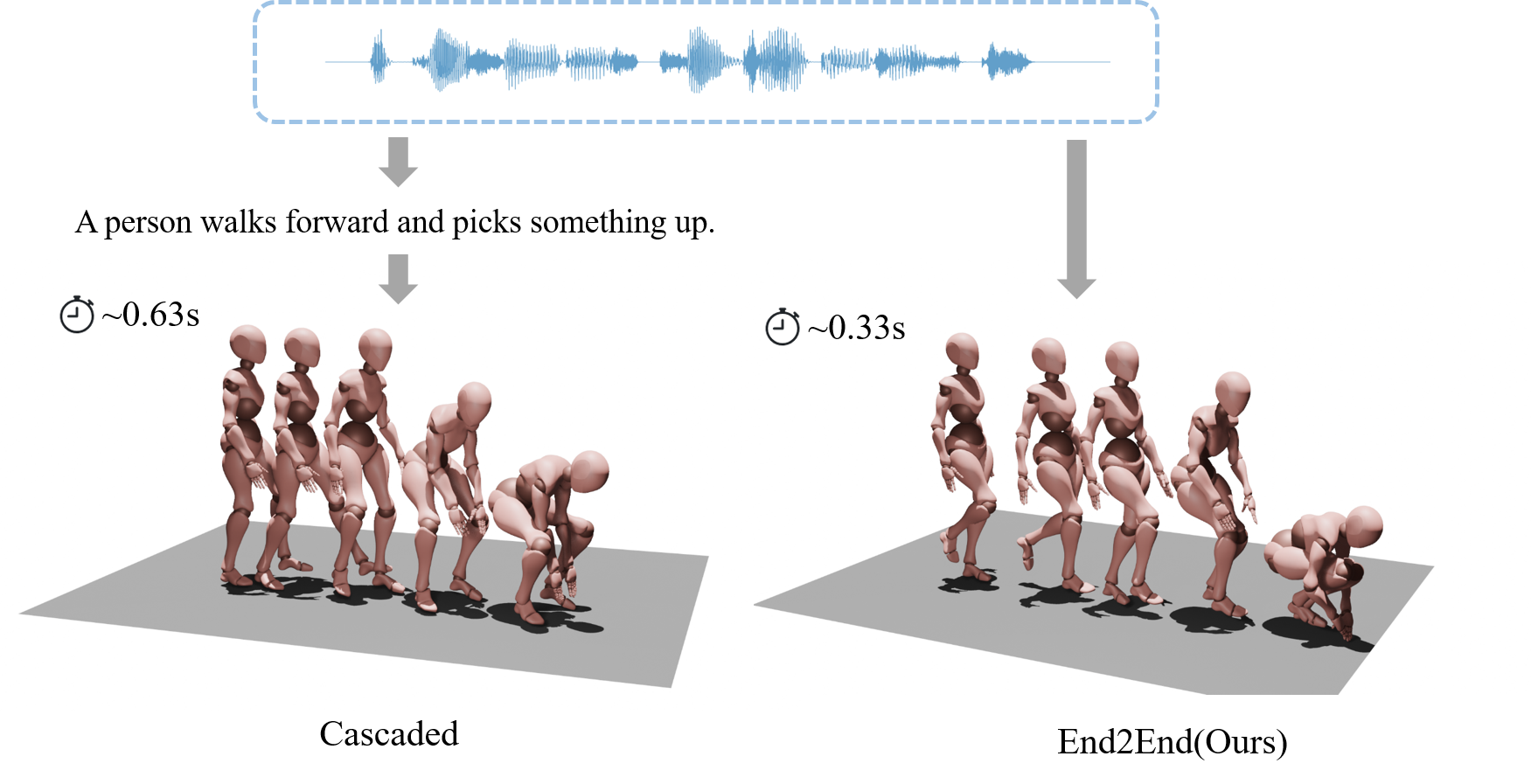}
    \caption{\textbf{Comparison with Cascaded Method.} Our end-to-end method can generate more than 50\% faster than cascaded approach, while maintaining high generation quality.
    }
    \vspace{-10pt}
    \label{fig:speed_performance}
\end{figure*}

% \textbf{Comparison to Cascaded Method.} The previous experiments have demonstrated that audio signals have equivalent semantic representation ability with text for human motion generation. In many real-world applications, a cascaded approach is commonly used, where audio is first converted into text before being processed by downstream tasks. However, when the semantic encoding of audio is sufficiently robust, our proposed end-to-end framework can achieve significantly higher efficiency than such cascaded methods. To validate this, we used the small version of Whisper~\cite{radford2023robust} for speech recognition. The output texts were then passed our text-based framework for motion generation, where texts are encoded with CLIP as input conditions. Performance testing was conducted on the same NVIDIA A6000 hardware. As shown in Tab.~\ref{tab:comparison_efficiency}, our end-to-end framework achieves over a 50\% improvement in generation efficiency compared to the two-stage cascaded approach, while maintaining high generation quality, as shown in Fig.~\ref{fig:speed_performance}. This highlights the practical advantages of bypassing text conversion in favor of direct audio input for motion generation.

\revise{\textbf{Comparison to Cascaded Method.} The previous experiments have demonstrated that audio signals have equivalent semantic representation ability with text for human motion generation. In many real-world applications, a cascaded approach is commonly used, where audio is first converted into text before being processed by downstream tasks. However, when the semantic encoding of audio is sufficiently robust, our proposed end-to-end framework can achieve significantly higher efficiency than such cascaded methods. To validate this, we used the small version of Whisper~\cite{radford2023robust} for speech recognition. The output texts were then passed our text-based framework for motion generation, where texts are encoded with CLIP as input conditions. 
Due to the high accuracy of speech recognition~\cite{radford2023robust}, the performance of motion sequences generated by the cascaded approach is generally comparable to those generated by text-to-motion. Therefore, we focus more on comparing the efficiency differences between the two approaches.
Performance testing was conducted on the same NVIDIA A6000 hardware. As shown in Tab.~\ref{tab:comparison_efficiency}, our end-to-end framework achieves over a 50\% improvement in generation efficiency compared to the two-stage cascaded approach, while maintaining high generation quality, as shown in Fig.~\ref{fig:speed_performance}. Specifically,  we obtained an average generation efficiency of approximately 360 frames per second. This highlights the practical advantages of bypassing text conversion in favor of direct audio input for motion generation.}

\begin{table}[h]
    \centering
    % \vspace{-10px}
    \renewcommand{\arraystretch}{1.1}
    \resizebox{\columnwidth}{!}{
    \begin{tabular}{l c c}
        \toprule
        \textbf{Framework} & \textbf{Pipeline} & \textbf{Speed}\small{ (sample/sec)} \\
        \hline
        End2End (Ours) & audio$\rightarrow$motion & 2.6951 \\
        Cascaded & audio$\rightarrow$text$\rightarrow$motion & 1.5093 \\
        \bottomrule
    \end{tabular}
    }
    \vspace{-10pt}
    \caption{\textbf{Comparison to Cascaded Method.} Our end-to-end framework achieves over a 50\% improvement in generation efficiency compared to the two-stage cascaded approach. The speed means the number of samples generated per second.}
    \label{tab:comparison_efficiency}
\end{table}

\begin{table}[htb]
    \centering
    \setlength{\tabcolsep}{1.2mm}
    \renewcommand{\arraystretch}{1.1}
    % \vspace{-2pt}
    \resizebox{\columnwidth}{!}{
    \begin{tabular}{lccccc}
        \toprule
        % \hline
        \textbf{Models} & \textbf{FID}$\downarrow$ & \textbf{R-top1}$\uparrow$ & \textbf{R-top2}$\uparrow$ & \textbf{R-top3}$\uparrow$ & \textbf{MM Dist}$\downarrow$ \\
        % \midrule
        \hline
        Encodec~\cite{borsos2023audiolm} & 2.920 & 0.233 & 0.370 & 0.464 & 5.548 \\
        ImageBind~\cite{girdhar2023imagebind} & 2.712 & 0.143 & 0.239 & 0.309 & 7.295 \\
        WavLM~\cite{chen2022wavlm} & \textbf{0.113} & \textbf{0.426} & \textbf{0.645} & \textbf{0.771} & \textbf{2.817} \\
        \bottomrule
        % \hline
    \end{tabular}
    }
    \vspace{-10pt}
    \caption{\textbf{Ablation on Pre-trained Audio Feature Extractors.} WavLM's strong generalization across multiple downstream tasks leads to better performance in our audio-instructed motion generation, emphasizing the importance of effective pre-trained models in enhancing motion synthesis quality.}
    \label{tab:comparison_audio_feature_extraction}
    \vspace{-10pt}
\end{table}

\begin{table}[h]
    \centering
    % \vspace{-10px}
    \setlength{\tabcolsep}{1mm}
    \renewcommand{\arraystretch}{1.1}
    \resizebox{\columnwidth}{!}{
    \begin{tabular}{l c c c c c}
        \toprule
        \textbf{Audio Encoding} & \textbf{FID}$\downarrow$ & \textbf{R-top1}$\uparrow$ & \textbf{R-top2}$\uparrow$ & \textbf{R-top3}$\uparrow$ & \textbf{MM Dist}$\downarrow$ \\
        \hline
        AvgPool-8  & 0.733 & 0.376  & 0.539  & 0.627  & 4.244   \\
        % \hline
        % AvgPool-16   & 0.779 & 0.431  & 0.598  & 0.699  & 3.647   \\
        % \hline
        AvgPool-32   & 1.258 & 0.406  & 0.581  & 0.677  & 3.706   \\
        % \hline
        AvgPool-64   & 0.571 & 0.408  & 0.591  & 0.686  & 3.640   \\
        % \hline
        Conv1d          & 2.000 & 0.342  & 0.503  & 0.602  & 4.471   \\
        % \hline
        Transformer     & 0.523 & 0.496  & 0.680  & 0.778  & 3.259   \\
        % \hline
        \textbf{Mem-Retr(Ours)}              & \textbf{0.121} & \textbf{0.519}  & \textbf{0.713}  & \textbf{0.808}  & \textbf{1.221} \\
        \bottomrule
    \end{tabular}
    }
    \vspace{-5px}
    \caption{\textbf{Ablation on Audio Feature Compression for Conditioning.} 
    % Compared with other architecture design, our proposed memory-retrieval based module shows a significant improvement.
    Our proposed memory-retrieval based module demonstrates a significant improvement over other architectural designs. This improvement highlights its effectiveness in compressing audio features as conditions, leading to better performance in motion generation tasks.
    }
    \vspace{-20px}
    \label{tab:comparison_audio_encoding}
\end{table}

% In previous sections (Section \ref{sec:introduction}), we discussed a two-staged baseline method for generating human actions from instructive audio. This approach first utilizes the Whisper~\cite{radford2023robust} speech recognition algorithm to convert audio into text, followed by a text-to-motion generation algorithm such as Momask~\cite{guo2024momask} to accomplish the human action generation task. To compare the efficiency differences between the two-stage baseline method and our proposed end-to-end algorithm, this study tests the inference efficiency using the test set of the audio-to-motion dataset synthesized from HumanML3D as input.

% Specifically, we employed the small version of Whisper for speech recognition, with the resulting output directly fed into Momask for inference. The performance testing was conducted on the same NVIDIA A6000 hardware. The results are presented in Table \ref{tab:comparison_efficiency}. As indicated in the table, the end-to-end generation algorithm proposed in this paper demonstrates over a 50\% improvement in generation efficiency compared to the two-stage baseline method.

\textbf{Pre-trained Audio Feature Extractors.} In audio research, numerous pre-trained models are available. To evaluate the effectiveness of our chosen WavLM model, we replaced it with alternative pre-trained models and conducted training and testing on the Original KIT-ML dataset. The results, shown in Tab.~\ref{tab:comparison_audio_feature_extraction}, demonstrate that WavLM significantly outperforms the alternatives. This improvement is likely due to WavLM’s strong generalization across multiple downstream tasks, whereas other models are often optimized for specific tasks.

\textbf{Audio Feature Compression for Conditioning.} As discussed in Sec.~\ref{subsec:memory_retrieval_based}, the extracted audio features exhibit significant variation in sequence length across different samples, making an effective compression strategy essential for subsequent audio-conditioned motion generation. We explored several methods for compressing sequence lengths, including Average Pooling (AvgPool) to fixed lengths (e.g., 8, 32, 64), Conv1d, and Transformer Encoder, for encoding audio conditions. The results, presented in Tab.~\ref{tab:comparison_audio_encoding}, are based on models trained and tested on the Original HumanML3D dataset. The results indicate that AvgPool delivers unsatisfactory performance, likely due to its lack of learnable parameters, which limits its adaptability to varying audio features. Similarly, Conv1d performs poorly, as its fixed receptive field struggles to capture long-range dependencies. The Transformer Encoder, although more capable, still appears to incorporate irrelevant audio content that is not directly related to motion description. In contrast, our proposed feature compression module effectively handles the significant variation in audio sequence lengths, producing more refined and semantics-aware audio conditions. This improvement leads to enhanced performance in generating accurate, audio-conditioned motions.

\section{Conclusion}
\label{sec:conclusion}

\revise{
In conclusion, this paper introduces a novel task: motion generation based on audio instructions. 
Compared to text-based interactions, audio offers a more convenient and natural mode of user interaction in real-world scenarios. 
We propose an end-to-end framework built on a masked generative transformer, enhanced with a memory-retrieval based attention module to effectively address the challenges of sparse and extended audio signals. This end-to-end paradigm demonstrates superior inference efficiency compared to the cascaded method. Moreover, it requires only a single deep learning model during inference, simplifying both deployment and training. Additionally, the human brain typically processes audio signals directly, without the need to convert them into text. Consequently, we believe that adopting a similar direct processing approach is more natural and effective.
Additionally, we augment existing text-based datasets by rephrasing textual descriptions into a more conversational, spoken language style and synthesizing corresponding audio with diverse speaker identities. 
Our experimental results demonstrate the effectiveness and efficiency of this framework in generating human motion directly from audio instructions. 
In parallel, we test the model using real recorded user voice commands as well as noisy voice command audio, demonstrating the robustness of the proposed model in real-world scenario.
} 

\reviseminor{
While our proposed framework demonstrates strong performance in generating human motion from English short audio instructions, several limitations remain to be addressed. First, the effectiveness of the model on multilingual instructions has not yet been explored. Second, although our current experiments focus on relatively short, single-stage instructions, the model’s performance on longer or multi-stage instructions requiring sequential actions remains untested. Investigating these aspects represents an important direction for future work, particularly in enhancing applicability in more diverse and complex real-world scenarios.
}

% For future work, we plan to explore the unique characteristics of instructive audio as conditioning signals, focusing on aspects that text cannot capture, such as pronunciation, intonation, and their influence on motion generation.

% \revise{For future work, we plan to explore the unique vocal features of instructive audio as conditioning signals, focusing on aspects that text cannot capture, such as pronunciation, intonation, and their influence on motion generation. Specifically, the capability of Tortoise~\cite{betker2023better} to generate voice commands with distinct vocal features could be leveraged to synthesize a more diverse range of audio instructions. These audio instructions and motion samples could then be annotated using methods such as sentiment analysis or clustering to create a differentiated dataset. Subsequently, this dataset could be utilized to train the model ability to leverage vocal features effectively.}

\reviseminor{
In future work, we intend to investigate the utilization of the significant potential present in instructional audio as conditioning signals, particularly focusing on elements beyond textual content, such as pronunciation, intonation, and their impact on motion generation. Specifically, the Tortoise model~\cite{betker2023better} offers the capability to generate voice commands exhibiting diverse vocal characteristics, which can be harnessed to produce a broader spectrum of audio instructions. These audio instructions, along with corresponding motion samples, can be annotated using techniques like sentiment analysis or clustering to construct a nuanced dataset. This dataset would subsequently facilitate the training of models adept at effectively leveraging vocal features.
}

\section*{Acknowledgments}
The authors would also like to thank the anonymous referees for their valuable comments and helpful suggestions. The work was partially supported by grants from Shenzhen Science and Technology Program (Grant No. JSGG20220831105002004).

\bibliographystyle{elsarticle-num-names} 
% \bibliography{cvm}

\begin{thebibliography}{57}
\expandafter\ifx\csname natexlab\endcsname\relax\def\natexlab#1{#1}\fi
\providecommand{\url}[1]{\texttt{#1}}
\providecommand{\href}[2]{#2}
\providecommand{\path}[1]{#1}
\providecommand{\DOIprefix}{doi:}
\providecommand{\ArXivprefix}{arXiv:}
\providecommand{\URLprefix}{URL: }
\providecommand{\Pubmedprefix}{pmid:}
\providecommand{\doi}[1]{\href{http://dx.doi.org/#1}{\path{#1}}}
\providecommand{\Pubmed}[1]{\href{pmid:#1}{\path{#1}}}
\providecommand{\bibinfo}[2]{#2}
\ifx\xfnm\relax \def\xfnm[#1]{\unskip,\space#1}\fi
%Type = Misc
\bibitem[{Authors(2012{\natexlab{a}})}]{Authors12}
\bibinfo{author}{Authors}, \bibinfo{title}{The frobnicatable foo filter}, \bibinfo{year}{2012}{\natexlab{a}}. \bibinfo{note}{Face and Gesture submission ID 324. Supplied as additional material {\tt fg324.pdf}}.
%Type = Misc
\bibitem[{Authors(2012{\natexlab{b}})}]{Authors12b}
\bibinfo{author}{Authors}, \bibinfo{title}{Frobnication tutorial}, \bibinfo{year}{2012}{\natexlab{b}}. \bibinfo{note}{Supplied as additional material {\tt tr.pdf}}.
%Type = Article
\bibitem[{Alpher(2002)}]{Alpher02}
\bibinfo{author}{A.~Alpher},
\newblock \bibinfo{title}{Frobnication},
\newblock \bibinfo{journal}{Journal of Foo} \bibinfo{volume}{12} (\bibinfo{year}{2002}) \bibinfo{pages}{234--778}.
%Type = Article
\bibitem[{Alpher and Fotheringham-Smythe(2003)}]{Alpher03}
\bibinfo{author}{A.~Alpher}, \bibinfo{author}{J.~P.~N. Fotheringham-Smythe},
\newblock \bibinfo{title}{Frobnication revisited},
\newblock \bibinfo{journal}{Journal of Foo} \bibinfo{volume}{13} (\bibinfo{year}{2003}) \bibinfo{pages}{234--778}.
%Type = Article
\bibitem[{Alpher et~al.(2004)Alpher, Fotheringham-Smythe, and Gamow}]{Alpher04}
\bibinfo{author}{A.~Alpher}, \bibinfo{author}{J.~P.~N. Fotheringham-Smythe}, \bibinfo{author}{G.~Gamow},
\newblock \bibinfo{title}{Can a machine frobnicate?},
\newblock \bibinfo{journal}{Journal of Foo} \bibinfo{volume}{14} (\bibinfo{year}{2004}) \bibinfo{pages}{234--778}.
%Type = Article
\bibitem[{Goodfellow et~al.(2020)Goodfellow, Pouget-Abadie, Mirza, Xu, Warde-Farley, Ozair, Courville, and Bengio}]{goodfellow_generative_2020}
\bibinfo{author}{I.~Goodfellow}, \bibinfo{author}{J.~Pouget-Abadie}, \bibinfo{author}{M.~Mirza}, \bibinfo{author}{B.~Xu}, \bibinfo{author}{D.~Warde-Farley}, \bibinfo{author}{S.~Ozair}, \bibinfo{author}{A.~Courville}, \bibinfo{author}{Y.~Bengio},
\newblock \bibinfo{title}{Generative adversarial networks},
\newblock \bibinfo{journal}{Communications of the ACM} \bibinfo{volume}{63} (\bibinfo{year}{2020}) \bibinfo{pages}{139--144}. \bibinfo{note}{Publisher: ACM New York, NY, USA}.
%Type = Article
\bibitem[{Kingma(2013)}]{kingma2013auto}
\bibinfo{author}{D.~P. Kingma},
\newblock \bibinfo{title}{Auto-encoding variational bayes},
\newblock \bibinfo{journal}{arXiv preprint arXiv:1312.6114}  (\bibinfo{year}{2013}).
%Type = Article
\bibitem[{Papamakarios et~al.(2021)Papamakarios, Nalisnick, Rezende, Mohamed, and Lakshminarayanan}]{papamakarios2021normalizing}
\bibinfo{author}{G.~Papamakarios}, \bibinfo{author}{E.~Nalisnick}, \bibinfo{author}{D.~J. Rezende}, \bibinfo{author}{S.~Mohamed}, \bibinfo{author}{B.~Lakshminarayanan},
\newblock \bibinfo{title}{Normalizing flows for probabilistic modeling and inference},
\newblock \bibinfo{journal}{Journal of Machine Learning Research} \bibinfo{volume}{22} (\bibinfo{year}{2021}) \bibinfo{pages}{1--64}.
%Type = Article
\bibitem[{Ho et~al.(2020)Ho, Jain, and Abbeel}]{ho2020denoising}
\bibinfo{author}{J.~Ho}, \bibinfo{author}{A.~Jain}, \bibinfo{author}{P.~Abbeel},
\newblock \bibinfo{title}{Denoising diffusion probabilistic models},
\newblock \bibinfo{journal}{Advances in neural information processing systems} \bibinfo{volume}{33} (\bibinfo{year}{2020}) \bibinfo{pages}{6840--6851}.
%Type = Inproceedings
\bibitem[{Chang et~al.(2022)Chang, Zhang, Jiang, Liu, and Freeman}]{chang2022maskgit}
\bibinfo{author}{H.~Chang}, \bibinfo{author}{H.~Zhang}, \bibinfo{author}{L.~Jiang}, \bibinfo{author}{C.~Liu}, \bibinfo{author}{W.~T. Freeman},
\newblock \bibinfo{title}{Maskgit: Masked generative image transformer},
\newblock in: \bibinfo{booktitle}{Proceedings of the IEEE Conference on Computer Vision and Pattern Recognition}, \bibinfo{year}{2022}, pp. \bibinfo{pages}{11315--11325}.
%Type = Article
\bibitem[{Chen et~al.(2022)Chen, Wang, Chen, Wu, Liu, Chen, Li, Kanda, Yoshioka, Xiao et~al.}]{chen2022wavlm}
\bibinfo{author}{S.~Chen}, \bibinfo{author}{C.~Wang}, \bibinfo{author}{Z.~Chen}, \bibinfo{author}{Y.~Wu}, \bibinfo{author}{S.~Liu}, \bibinfo{author}{Z.~Chen}, \bibinfo{author}{J.~Li}, \bibinfo{author}{N.~Kanda}, \bibinfo{author}{T.~Yoshioka}, \bibinfo{author}{X.~Xiao}, et~al.,
\newblock \bibinfo{title}{Wavlm: Large-scale self-supervised pre-training for full stack speech processing},
\newblock \bibinfo{journal}{IEEE Journal of Selected Topics in Signal Processing} \bibinfo{volume}{16} (\bibinfo{year}{2022}) \bibinfo{pages}{1505--1518}.
%Type = Article
\bibitem[{D{\'e}fossez et~al.(2022)D{\'e}fossez, Copet, Synnaeve, and Adi}]{defossez2022high}
\bibinfo{author}{A.~D{\'e}fossez}, \bibinfo{author}{J.~Copet}, \bibinfo{author}{G.~Synnaeve}, \bibinfo{author}{Y.~Adi},
\newblock \bibinfo{title}{High fidelity neural audio compression},
\newblock \bibinfo{journal}{arXiv preprint arXiv:2210.13438}  (\bibinfo{year}{2022}).
%Type = Inproceedings
\bibitem[{Radford et~al.(2023)Radford, Kim, Xu, Brockman, McLeavey, and Sutskever}]{radford2023robust}
\bibinfo{author}{A.~Radford}, \bibinfo{author}{J.~W. Kim}, \bibinfo{author}{T.~Xu}, \bibinfo{author}{G.~Brockman}, \bibinfo{author}{C.~McLeavey}, \bibinfo{author}{I.~Sutskever},
\newblock \bibinfo{title}{Robust speech recognition via large-scale weak supervision},
\newblock in: \bibinfo{booktitle}{International conference on machine learning}, \bibinfo{organization}{PMLR}, \bibinfo{year}{2023}, pp. \bibinfo{pages}{28492--28518}.
%Type = Inproceedings
\bibitem[{Girdhar et~al.(2023)Girdhar, El-Nouby, Liu, Singh, Alwala, Joulin, and Misra}]{girdhar2023imagebind}
\bibinfo{author}{R.~Girdhar}, \bibinfo{author}{A.~El-Nouby}, \bibinfo{author}{Z.~Liu}, \bibinfo{author}{M.~Singh}, \bibinfo{author}{K.~V. Alwala}, \bibinfo{author}{A.~Joulin}, \bibinfo{author}{I.~Misra},
\newblock \bibinfo{title}{Imagebind: One embedding space to bind them all},
\newblock in: \bibinfo{booktitle}{Proceedings of the IEEE Conference on Computer Vision and Pattern Recognition}, \bibinfo{year}{2023}, pp. \bibinfo{pages}{15180--15190}.
%Type = Inproceedings
\bibitem[{Ahuja and Morency(2019)}]{ahuja2019language2pose}
\bibinfo{author}{C.~Ahuja}, \bibinfo{author}{L.-P. Morency},
\newblock \bibinfo{title}{Language2pose: Natural language grounded pose forecasting},
\newblock in: \bibinfo{booktitle}{2019 International Conference on 3D Vision}, \bibinfo{organization}{IEEE}, \bibinfo{year}{2019}, pp. \bibinfo{pages}{719--728}.
%Type = Inproceedings
\bibitem[{Ghosh et~al.(2021)Ghosh, Cheema, Oguz, Theobalt, and Slusallek}]{ghosh2021synthesis}
\bibinfo{author}{A.~Ghosh}, \bibinfo{author}{N.~Cheema}, \bibinfo{author}{C.~Oguz}, \bibinfo{author}{C.~Theobalt}, \bibinfo{author}{P.~Slusallek},
\newblock \bibinfo{title}{Synthesis of compositional animations from textual descriptions},
\newblock in: \bibinfo{booktitle}{Proceedings of the IEEE international conference on computer vision}, \bibinfo{year}{2021}, pp. \bibinfo{pages}{1396--1406}.
%Type = Article
\bibitem[{Lin et~al.(2018)Lin, Wu, Corona, Tai, Huang, and Mooney}]{lin2018generating}
\bibinfo{author}{A.~S. Lin}, \bibinfo{author}{L.~Wu}, \bibinfo{author}{R.~Corona}, \bibinfo{author}{K.~Tai}, \bibinfo{author}{Q.~Huang}, \bibinfo{author}{R.~J. Mooney},
\newblock \bibinfo{title}{Generating animated videos of human activities from natural language descriptions},
\newblock \bibinfo{journal}{Learning} \bibinfo{volume}{1} (\bibinfo{year}{2018}) \bibinfo{pages}{1}.
%Type = Inproceedings
\bibitem[{Tevet et~al.(2022)Tevet, Gordon, Hertz, Bermano, and Cohen-Or}]{tevet2022motionclip}
\bibinfo{author}{G.~Tevet}, \bibinfo{author}{B.~Gordon}, \bibinfo{author}{A.~Hertz}, \bibinfo{author}{A.~H. Bermano}, \bibinfo{author}{D.~Cohen-Or},
\newblock \bibinfo{title}{Motionclip: Exposing human motion generation to clip space},
\newblock in: \bibinfo{booktitle}{European Conference on Computer Vision}, \bibinfo{organization}{Springer}, \bibinfo{year}{2022}, pp. \bibinfo{pages}{358--374}.
%Type = Inproceedings
\bibitem[{Radford et~al.(2021)Radford, Kim, Hallacy, Ramesh, Goh, Agarwal, Sastry, Askell, Mishkin, Clark et~al.}]{radford2021learning}
\bibinfo{author}{A.~Radford}, \bibinfo{author}{J.~W. Kim}, \bibinfo{author}{C.~Hallacy}, \bibinfo{author}{A.~Ramesh}, \bibinfo{author}{G.~Goh}, \bibinfo{author}{S.~Agarwal}, \bibinfo{author}{G.~Sastry}, \bibinfo{author}{A.~Askell}, \bibinfo{author}{P.~Mishkin}, \bibinfo{author}{J.~Clark}, et~al.,
\newblock \bibinfo{title}{Learning transferable visual models from natural language supervision},
\newblock in: \bibinfo{booktitle}{International conference on machine learning}, \bibinfo{organization}{PMLR}, \bibinfo{year}{2021}, pp. \bibinfo{pages}{8748--8763}.
%Type = Inproceedings
\bibitem[{Petrovich et~al.(2022)Petrovich, Black, and Varol}]{petrovich2022temos}
\bibinfo{author}{M.~Petrovich}, \bibinfo{author}{M.~J. Black}, \bibinfo{author}{G.~Varol},
\newblock \bibinfo{title}{Temos: Generating diverse human motions from textual descriptions},
\newblock in: \bibinfo{booktitle}{European Conference on Computer Vision}, \bibinfo{organization}{Springer}, \bibinfo{year}{2022}, pp. \bibinfo{pages}{480--497}.
%Type = Inproceedings
\bibitem[{Guo et~al.(2022)Guo, Zou, Zuo, Wang, Ji, Li, and Cheng}]{guo2022generating}
\bibinfo{author}{C.~Guo}, \bibinfo{author}{S.~Zou}, \bibinfo{author}{X.~Zuo}, \bibinfo{author}{S.~Wang}, \bibinfo{author}{W.~Ji}, \bibinfo{author}{X.~Li}, \bibinfo{author}{L.~Cheng},
\newblock \bibinfo{title}{Generating diverse and natural 3d human motions from text},
\newblock in: \bibinfo{booktitle}{Proceedings of the IEEE Conference on Computer Vision and Pattern Recognition}, \bibinfo{year}{2022}, pp. \bibinfo{pages}{5152--5161}.
%Type = Article
\bibitem[{Van Den~Oord et~al.(2017)Van Den~Oord, Vinyals et~al.}]{van2017neural}
\bibinfo{author}{A.~Van Den~Oord}, \bibinfo{author}{O.~Vinyals}, et~al.,
\newblock \bibinfo{title}{Neural discrete representation learning},
\newblock \bibinfo{journal}{Advances in neural information processing systems} \bibinfo{volume}{30} (\bibinfo{year}{2017}).
%Type = Inproceedings
\bibitem[{Zhang et~al.(2023)Zhang, Zhang, Cun, Zhang, Zhao, Lu, Shen, and Shan}]{zhang2023generating}
\bibinfo{author}{J.~Zhang}, \bibinfo{author}{Y.~Zhang}, \bibinfo{author}{X.~Cun}, \bibinfo{author}{Y.~Zhang}, \bibinfo{author}{H.~Zhao}, \bibinfo{author}{H.~Lu}, \bibinfo{author}{X.~Shen}, \bibinfo{author}{Y.~Shan},
\newblock \bibinfo{title}{Generating human motion from textual descriptions with discrete representations},
\newblock in: \bibinfo{booktitle}{Proceedings of the IEEE conference on computer vision and pattern recognition}, \bibinfo{year}{2023}, pp. \bibinfo{pages}{14730--14740}.
%Type = Inproceedings
\bibitem[{Kim et~al.(2023)Kim, Kim, and Choi}]{kim2023flame}
\bibinfo{author}{J.~Kim}, \bibinfo{author}{J.~Kim}, \bibinfo{author}{S.~Choi},
\newblock \bibinfo{title}{Flame: Free-form language-based motion synthesis \& editing},
\newblock in: \bibinfo{booktitle}{Proceedings of the AAAI Conference on Artificial Intelligence}, volume~\bibinfo{volume}{37}, \bibinfo{year}{2023}, pp. \bibinfo{pages}{8255--8263}.
%Type = Inproceedings
\bibitem[{Yuan et~al.(2023)Yuan, Song, Iqbal, Vahdat, and Kautz}]{yuan2023physdiff}
\bibinfo{author}{Y.~Yuan}, \bibinfo{author}{J.~Song}, \bibinfo{author}{U.~Iqbal}, \bibinfo{author}{A.~Vahdat}, \bibinfo{author}{J.~Kautz},
\newblock \bibinfo{title}{Physdiff: Physics-guided human motion diffusion model},
\newblock in: \bibinfo{booktitle}{Proceedings of the IEEE international conference on computer vision}, \bibinfo{year}{2023}, pp. \bibinfo{pages}{16010--16021}.
%Type = Inproceedings
\bibitem[{Tevet et~al.(2023)Tevet, Raab, Gordon, Shafir, Cohen-or, and Bermano}]{tevet2023human}
\bibinfo{author}{G.~Tevet}, \bibinfo{author}{S.~Raab}, \bibinfo{author}{B.~Gordon}, \bibinfo{author}{Y.~Shafir}, \bibinfo{author}{D.~Cohen-or}, \bibinfo{author}{A.~H. Bermano},
\newblock \bibinfo{title}{Human motion diffusion model},
\newblock in: \bibinfo{booktitle}{The Eleventh International Conference on Learning Representations}, \bibinfo{year}{2023}.
%Type = Inproceedings
\bibitem[{Guo et~al.(2024)Guo, Mu, Javed, Wang, and Cheng}]{guo2024momask}
\bibinfo{author}{C.~Guo}, \bibinfo{author}{Y.~Mu}, \bibinfo{author}{M.~G. Javed}, \bibinfo{author}{S.~Wang}, \bibinfo{author}{L.~Cheng},
\newblock \bibinfo{title}{Momask: Generative masked modeling of 3d human motions},
\newblock in: \bibinfo{booktitle}{Proceedings of the IEEE Conference on Computer Vision and Pattern Recognition}, \bibinfo{year}{2024}, pp. \bibinfo{pages}{1900--1910}.
%Type = Inproceedings
\bibitem[{Tang et~al.(2018)Tang, Jia, and Mao}]{tang2018dance}
\bibinfo{author}{T.~Tang}, \bibinfo{author}{J.~Jia}, \bibinfo{author}{H.~Mao},
\newblock \bibinfo{title}{Dance with melody: An lstm-autoencoder approach to music-oriented dance synthesis},
\newblock in: \bibinfo{booktitle}{Proceedings of the 26th ACM international conference on Multimedia}, \bibinfo{year}{2018}, pp. \bibinfo{pages}{1598--1606}.
%Type = Article
\bibitem[{Graves and Graves(2012)}]{graves2012long}
\bibinfo{author}{A.~Graves}, \bibinfo{author}{A.~Graves},
\newblock \bibinfo{title}{Long short-term memory},
\newblock \bibinfo{journal}{Supervised sequence labelling with recurrent neural networks}  (\bibinfo{year}{2012}) \bibinfo{pages}{37--45}.
%Type = Article
\bibitem[{Lee et~al.(2019)Lee, Yang, Liu, Wang, Lu, Yang, and Kautz}]{lee2019dancing}
\bibinfo{author}{H.-Y. Lee}, \bibinfo{author}{X.~Yang}, \bibinfo{author}{M.-Y. Liu}, \bibinfo{author}{T.-C. Wang}, \bibinfo{author}{Y.-D. Lu}, \bibinfo{author}{M.-H. Yang}, \bibinfo{author}{J.~Kautz},
\newblock \bibinfo{title}{Dancing to music},
\newblock \bibinfo{journal}{Advances in neural information processing systems} \bibinfo{volume}{32} (\bibinfo{year}{2019}).
%Type = Article
\bibitem[{Chen et~al.(2021)Chen, Tan, Lei, Zhang, Guo, Zhang, and Hu}]{chen2021choreomaster}
\bibinfo{author}{K.~Chen}, \bibinfo{author}{Z.~Tan}, \bibinfo{author}{J.~Lei}, \bibinfo{author}{S.-H. Zhang}, \bibinfo{author}{Y.-C. Guo}, \bibinfo{author}{W.~Zhang}, \bibinfo{author}{S.-M. Hu},
\newblock \bibinfo{title}{Choreomaster: choreography-oriented music-driven dance synthesis},
\newblock \bibinfo{journal}{ACM Transactions on Graphics} \bibinfo{volume}{40} (\bibinfo{year}{2021}) \bibinfo{pages}{1--13}.
%Type = Inproceedings
\bibitem[{Ginosar et~al.(2019)Ginosar, Bar, Kohavi, Chan, Owens, and Malik}]{ginosar2019learning}
\bibinfo{author}{S.~Ginosar}, \bibinfo{author}{A.~Bar}, \bibinfo{author}{G.~Kohavi}, \bibinfo{author}{C.~Chan}, \bibinfo{author}{A.~Owens}, \bibinfo{author}{J.~Malik},
\newblock \bibinfo{title}{Learning individual styles of conversational gesture},
\newblock in: \bibinfo{booktitle}{Proceedings of the IEEE Conference on Computer Vision and Pattern Recognition}, \bibinfo{year}{2019}, pp. \bibinfo{pages}{3497--3506}.
%Type = Inproceedings
\bibitem[{Kucherenko et~al.(2019)Kucherenko, Hasegawa, Henter, Kaneko, and Kjellstr{\"o}m}]{kucherenko2019analyzing}
\bibinfo{author}{T.~Kucherenko}, \bibinfo{author}{D.~Hasegawa}, \bibinfo{author}{G.~E. Henter}, \bibinfo{author}{N.~Kaneko}, \bibinfo{author}{H.~Kjellstr{\"o}m},
\newblock \bibinfo{title}{Analyzing input and output representations for speech-driven gesture generation},
\newblock in: \bibinfo{booktitle}{Proceedings of the 19th ACM International Conference on Intelligent Virtual Agents}, \bibinfo{year}{2019}, pp. \bibinfo{pages}{97--104}.
%Type = Inproceedings
\bibitem[{Li et~al.(2021)Li, Kang, Pei, Zhe, Zhang, He, and Bao}]{li2021audio2gestures}
\bibinfo{author}{J.~Li}, \bibinfo{author}{D.~Kang}, \bibinfo{author}{W.~Pei}, \bibinfo{author}{X.~Zhe}, \bibinfo{author}{Y.~Zhang}, \bibinfo{author}{Z.~He}, \bibinfo{author}{L.~Bao},
\newblock \bibinfo{title}{Audio2gestures: Generating diverse gestures from speech audio with conditional variational autoencoders},
\newblock in: \bibinfo{booktitle}{Proceedings of the IEEE International Conference on Computer Vision}, \bibinfo{year}{2021}, pp. \bibinfo{pages}{11293--11302}.
%Type = Article
\bibitem[{Zhu et~al.(2023)Zhu, Ma, Ro, Ci, Zhang, Shi, Gao, Tian, and Wang}]{zhu2023survey}
\bibinfo{author}{W.~Zhu}, \bibinfo{author}{X.~Ma}, \bibinfo{author}{D.~Ro}, \bibinfo{author}{H.~Ci}, \bibinfo{author}{J.~Zhang}, \bibinfo{author}{J.~Shi}, \bibinfo{author}{F.~Gao}, \bibinfo{author}{Q.~Tian}, \bibinfo{author}{Y.~Wang},
\newblock \bibinfo{title}{Human motion generation: A survey},
\newblock \bibinfo{journal}{IEEE Transactions on Pattern Analysis and Machine Intelligence}  (\bibinfo{year}{2023}).
%Type = Article
\bibitem[{Borsos et~al.(2023)Borsos, Marinier, Vincent, Kharitonov, Pietquin, Sharifi, Roblek, Teboul, Grangier, Tagliasacchi et~al.}]{borsos2023audiolm}
\bibinfo{author}{Z.~Borsos}, \bibinfo{author}{R.~Marinier}, \bibinfo{author}{D.~Vincent}, \bibinfo{author}{E.~Kharitonov}, \bibinfo{author}{O.~Pietquin}, \bibinfo{author}{M.~Sharifi}, \bibinfo{author}{D.~Roblek}, \bibinfo{author}{O.~Teboul}, \bibinfo{author}{D.~Grangier}, \bibinfo{author}{M.~Tagliasacchi}, et~al.,
\newblock \bibinfo{title}{Audiolm: a language modeling approach to audio generation},
\newblock \bibinfo{journal}{IEEE transactions on audio, speech, and language processing} \bibinfo{volume}{31} (\bibinfo{year}{2023}) \bibinfo{pages}{2523--2533}.
%Type = Article
\bibitem[{OpenAI et~al.(2024)OpenAI, Adler, Agarwal, Ahmad, Akkaya, Aleman, Almeida, Altenschmidt, Altman, Anadkat et~al.}]{openai2024gpt}
\bibinfo{author}{A.~J. OpenAI}, \bibinfo{author}{S.~Adler}, \bibinfo{author}{S.~Agarwal}, \bibinfo{author}{L.~Ahmad}, \bibinfo{author}{I.~Akkaya}, \bibinfo{author}{F.~L. Aleman}, \bibinfo{author}{D.~Almeida}, \bibinfo{author}{J.~Altenschmidt}, \bibinfo{author}{S.~Altman}, \bibinfo{author}{S.~Anadkat}, et~al.,
\newblock \bibinfo{title}{Gpt-4 technical report. 2023},
\newblock \bibinfo{journal}{URL: https://arxiv. org/abs/2303.08774}  (\bibinfo{year}{2024}).
%Type = Article
\bibitem[{Betker(2023)}]{betker2023better}
\bibinfo{author}{J.~Betker},
\newblock \bibinfo{title}{Better speech synthesis through scaling},
\newblock \bibinfo{journal}{arXiv preprint arXiv:2305.07243}  (\bibinfo{year}{2023}).
%Type = Article
\bibitem[{Plappert et~al.(2016)Plappert, Mandery, and Asfour}]{plappert2016kit}
\bibinfo{author}{M.~Plappert}, \bibinfo{author}{C.~Mandery}, \bibinfo{author}{T.~Asfour},
\newblock \bibinfo{title}{The kit motion-language dataset},
\newblock \bibinfo{journal}{Big data} \bibinfo{volume}{4} (\bibinfo{year}{2016}) \bibinfo{pages}{236--252}.
%Type = Article
\bibitem[{Vaswani(2017)}]{vaswani2017attention}
\bibinfo{author}{A.~Vaswani},
\newblock \bibinfo{title}{Attention is all you need},
\newblock \bibinfo{journal}{Advances in Neural Information Processing Systems}  (\bibinfo{year}{2017}).
%Type = Article
\bibitem[{Jiang et~al.(2024)Jiang, Liang, Yang, Lin, Zhong, and Zheng}]{jiang2024loopy}
\bibinfo{author}{J.~Jiang}, \bibinfo{author}{C.~Liang}, \bibinfo{author}{J.~Yang}, \bibinfo{author}{G.~Lin}, \bibinfo{author}{T.~Zhong}, \bibinfo{author}{Y.~Zheng},
\newblock \bibinfo{title}{Loopy: Taming audio-driven portrait avatar with long-term motion dependency},
\newblock \bibinfo{journal}{arXiv preprint arXiv:2409.02634}  (\bibinfo{year}{2024}).
%Type = Article
\bibitem[{Zhang et~al.(2022)Zhang, Cai, Pan, Hong, Guo, Yang, and Liu}]{zhang2022motiondiffuse}
\bibinfo{author}{M.~Zhang}, \bibinfo{author}{Z.~Cai}, \bibinfo{author}{L.~Pan}, \bibinfo{author}{F.~Hong}, \bibinfo{author}{X.~Guo}, \bibinfo{author}{L.~Yang}, \bibinfo{author}{Z.~Liu},
\newblock \bibinfo{title}{Motiondiffuse: Text-driven human motion generation with diffusion model},
\newblock \bibinfo{journal}{arXiv preprint arXiv:2208.15001}  (\bibinfo{year}{2022}).
%Type = Inproceedings
\bibitem[{Guo et~al.(2022)Guo, Zuo, Wang, and Cheng}]{guo2022tm2t}
\bibinfo{author}{C.~Guo}, \bibinfo{author}{X.~Zuo}, \bibinfo{author}{S.~Wang}, \bibinfo{author}{L.~Cheng},
\newblock \bibinfo{title}{Tm2t: Stochastic and tokenized modeling for the reciprocal generation of 3d human motions and texts},
\newblock in: \bibinfo{booktitle}{European Conference on Computer Vision}, \bibinfo{organization}{Springer}, \bibinfo{year}{2022}, pp. \bibinfo{pages}{580--597}.
%Type = Inproceedings
\bibitem[{Chen et~al.(2023)Chen, Jiang, Liu, Huang, Fu, Chen, and Yu}]{chen2023executing}
\bibinfo{author}{X.~Chen}, \bibinfo{author}{B.~Jiang}, \bibinfo{author}{W.~Liu}, \bibinfo{author}{Z.~Huang}, \bibinfo{author}{B.~Fu}, \bibinfo{author}{T.~Chen}, \bibinfo{author}{G.~Yu},
\newblock \bibinfo{title}{Executing your commands via motion diffusion in latent space},
\newblock in: \bibinfo{booktitle}{Proceedings of the IEEE Conference on Computer Vision and Pattern Recognition}, \bibinfo{year}{2023}, pp. \bibinfo{pages}{18000--18010}.
%Type = Inproceedings
\bibitem[{Zhang et~al.(2023)Zhang, Guo, Pan, Cai, Hong, Li, Yang, and Liu}]{zhang2023remodiffuse}
\bibinfo{author}{M.~Zhang}, \bibinfo{author}{X.~Guo}, \bibinfo{author}{L.~Pan}, \bibinfo{author}{Z.~Cai}, \bibinfo{author}{F.~Hong}, \bibinfo{author}{H.~Li}, \bibinfo{author}{L.~Yang}, \bibinfo{author}{Z.~Liu},
\newblock \bibinfo{title}{Remodiffuse: Retrieval-augmented motion diffusion model},
\newblock in: \bibinfo{booktitle}{Proceedings of the IEEE International Conference on Computer Vision}, \bibinfo{year}{2023}, pp. \bibinfo{pages}{364--373}.
%Type = Article
\bibitem[{Ao et~al.(2023)Ao, Zhang, and Liu}]{ao2023gesturediffuclip}
\bibinfo{author}{T.~Ao}, \bibinfo{author}{Z.~Zhang}, \bibinfo{author}{L.~Liu},
\newblock \bibinfo{title}{Gesturediffuclip: Gesture diffusion model with clip latents},
\newblock \bibinfo{journal}{ACM Transactions on Graphics} \bibinfo{volume}{42} (\bibinfo{year}{2023}) \bibinfo{pages}{1--18}.
%Type = Article
\bibitem[{Ao et~al.(2022)Ao, Gao, Lou, Chen, and Liu}]{ao2022rhythmic}
\bibinfo{author}{T.~Ao}, \bibinfo{author}{Q.~Gao}, \bibinfo{author}{Y.~Lou}, \bibinfo{author}{B.~Chen}, \bibinfo{author}{L.~Liu},
\newblock \bibinfo{title}{Rhythmic gesticulator: Rhythm-aware co-speech gesture synthesis with hierarchical neural embeddings},
\newblock \bibinfo{journal}{ACM Transactions on Graphics} \bibinfo{volume}{41} (\bibinfo{year}{2022}) \bibinfo{pages}{1--19}.
%Type = Inproceedings
\bibitem[{Dabral et~al.(2023)Dabral, Mughal, Golyanik, and Theobalt}]{dabral2023mofusion}
\bibinfo{author}{R.~Dabral}, \bibinfo{author}{M.~H. Mughal}, \bibinfo{author}{V.~Golyanik}, \bibinfo{author}{C.~Theobalt},
\newblock \bibinfo{title}{Mofusion: A framework for denoising-diffusion-based motion synthesis},
\newblock in: \bibinfo{booktitle}{Proceedings of the IEEE conference on computer vision and pattern recognition}, \bibinfo{year}{2023}, pp. \bibinfo{pages}{9760--9770}.
%Type = Inproceedings
\bibitem[{Gong et~al.(2023)Gong, Lian, Chang, Guo, Jiang, Zuo, Mi, and Wang}]{gong2023tm2d}
\bibinfo{author}{K.~Gong}, \bibinfo{author}{D.~Lian}, \bibinfo{author}{H.~Chang}, \bibinfo{author}{C.~Guo}, \bibinfo{author}{Z.~Jiang}, \bibinfo{author}{X.~Zuo}, \bibinfo{author}{M.~B. Mi}, \bibinfo{author}{X.~Wang},
\newblock \bibinfo{title}{Tm2d: Bimodality driven 3d dance generation via music-text integration},
\newblock in: \bibinfo{booktitle}{Proceedings of the IEEE International Conference on Computer Vision}, \bibinfo{year}{2023}, pp. \bibinfo{pages}{9942--9952}.
%Type = Inproceedings
\bibitem[{Li et~al.(2021)Li, Yang, Ross, and Kanazawa}]{li2021ai}
\bibinfo{author}{R.~Li}, \bibinfo{author}{S.~Yang}, \bibinfo{author}{D.~A. Ross}, \bibinfo{author}{A.~Kanazawa},
\newblock \bibinfo{title}{Ai choreographer: Music conditioned 3d dance generation with aist++},
\newblock in: \bibinfo{booktitle}{Proceedings of the IEEE International Conference on Computer Vision}, \bibinfo{year}{2021}, pp. \bibinfo{pages}{13401--13412}.
%Type = Article
\bibitem[{Zhang et~al.(2024)Zhang, Ao, Zhang, Gao, Lin, Chen, and Liu}]{zhang2024semantic}
\bibinfo{author}{Z.~Zhang}, \bibinfo{author}{T.~Ao}, \bibinfo{author}{Y.~Zhang}, \bibinfo{author}{Q.~Gao}, \bibinfo{author}{C.~Lin}, \bibinfo{author}{B.~Chen}, \bibinfo{author}{L.~Liu},
\newblock \bibinfo{title}{Semantic gesticulator: Semantics-aware co-speech gesture synthesis},
\newblock \bibinfo{journal}{ACM Transactions on Graphics} \bibinfo{volume}{43} (\bibinfo{year}{2024}) \bibinfo{pages}{1--17}.
%Type = Article
\bibitem[{Devlin(2018)}]{devlin2018bert}
\bibinfo{author}{J.~Devlin},
\newblock \bibinfo{title}{Bert: Pre-training of deep bidirectional transformers for language understanding},
\newblock \bibinfo{journal}{arXiv preprint arXiv:1810.04805}  (\bibinfo{year}{2018}).
%Type = Article
\bibitem[{Chang et~al.(2023)Chang, Zhang, Barber, Maschinot, Lezama, Jiang, Yang, Murphy, Freeman, Rubinstein et~al.}]{chang2023muse}
\bibinfo{author}{H.~Chang}, \bibinfo{author}{H.~Zhang}, \bibinfo{author}{J.~Barber}, \bibinfo{author}{A.~Maschinot}, \bibinfo{author}{J.~Lezama}, \bibinfo{author}{L.~Jiang}, \bibinfo{author}{M.-H. Yang}, \bibinfo{author}{K.~Murphy}, \bibinfo{author}{W.~T. Freeman}, \bibinfo{author}{M.~Rubinstein}, et~al.,
\newblock \bibinfo{title}{Muse: Text-to-image generation via masked generative transformers},
\newblock \bibinfo{journal}{arXiv preprint arXiv:2301.00704}  (\bibinfo{year}{2023}).
%Type = Article
\bibitem[{Liu et~al.(2025)Liu, Zhang, Kim, Garrido, Shapiro, and Olszewski}]{liu2025contextual}
\bibinfo{author}{P.~Liu}, \bibinfo{author}{P.~Zhang}, \bibinfo{author}{H.~Kim}, \bibinfo{author}{P.~Garrido}, \bibinfo{author}{A.~Shapiro}, \bibinfo{author}{K.~Olszewski},
\newblock \bibinfo{title}{Contextual gesture: Co-speech gesture video generation through context-aware gesture representation},
\newblock \bibinfo{journal}{arXiv preprint arXiv:2502.07239}  (\bibinfo{year}{2025}).
%Type = Article
\bibitem[{Zhou et~al.(2023)Zhou, Li, Zeng, Aristidou, Zhang, Chen, and Tu}]{zhou2023lets}
\bibinfo{author}{Q.~Zhou}, \bibinfo{author}{M.~Li}, \bibinfo{author}{Q.~Zeng}, \bibinfo{author}{A.~Aristidou}, \bibinfo{author}{X.~Zhang}, \bibinfo{author}{L.~Chen}, \bibinfo{author}{C.~Tu},
\newblock \bibinfo{title}{Let’s all dance: Enhancing amateur dance motions},
\newblock \bibinfo{journal}{Computational Visual Media}  (\bibinfo{year}{2023}).
%Type = Inproceedings
\bibitem[{Liu et~al.(2022)Liu, Xu, Wu, Zhou, Wu, and Zhou}]{liu2022semantic}
\bibinfo{author}{X.~Liu}, \bibinfo{author}{Y.~Xu}, \bibinfo{author}{Q.~Wu}, \bibinfo{author}{H.~Zhou}, \bibinfo{author}{W.~Wu}, \bibinfo{author}{B.~Zhou},
\newblock \bibinfo{title}{Semantic-aware implicit neural audio-driven video portrait generation},
\newblock in: \bibinfo{booktitle}{Proceedings of the European Conference on Computer Vision (ECCV)}, \bibinfo{year}{2022}, pp. \bibinfo{pages}{112--129}.
%Type = Article
\bibitem[{Shi et~al.(2024)Shi, Feng, Gao, and Zhu}]{shi2024generating}
\bibinfo{author}{M.~Shi}, \bibinfo{author}{W.~Feng}, \bibinfo{author}{L.~Gao}, \bibinfo{author}{D.~Zhu},
\newblock \bibinfo{title}{Generating diverse clothed 3d human animations via a generative model},
\newblock \bibinfo{journal}{Computational Visual Media} \bibinfo{volume}{10} (\bibinfo{year}{2024}) \bibinfo{pages}{261--277}.

\end{thebibliography}

\end{document}